\documentclass[useams,usenatbib]{mn2e}
\pdfoutput=1
\usepackage{graphicx}
\usepackage{mathptmx}
\usepackage{pdfsync}
\usepackage{amsmath}
\usepackage{amssymb}
\usepackage{aas_macros}
\usepackage[usenames,dvipsnames]{color}
\usepackage[pdfpagelabels]{hyperref}
\hypersetup{pdfauthor={Michael J. Williams, Martin Bureau and Michele
Cappellari},pdftitle={The Tully-Fisher relations of early-type spiral
and S0 galaxies},breaklinks=true}

\usepackage[a4paper,total={17.8cm,24.0cm},centering]{geometry}

\newcommand{\refsec}[1]{Section~\ref{#1}}
\newcommand{\reffig}[1]{Fig.~\ref{#1}}
\newcommand{\refeq}[1]{equation~(\ref{#1})}
\newcommand{\kms}{km\,s$^{-1}$}
\newcommand{\ud}{\mathrm{d}}

\newcommand{\reff}{\hbox{\ensuremath{R_\mathrm{e}}}}
\newcommand{\hi}{\hbox{H\,\textsc{i}}}
\newcommand{\vhi}{\ensuremath{v_\mathrm{H\,\textsc{i}}}}
\newcommand{\vdrift}{\ensuremath{v_\mathrm{drift}}}
\newcommand{\vmodel}{\ensuremath{v_\mathrm{model}}}
\newcommand{\vgas}{\ensuremath{v_\mathrm{gas}}}
\newcommand{\vc}{\ensuremath{v_\mathrm{c}}}
\newcommand{\halpha}{H$\alpha$}
\defcitealias{Bedregal:2006}{BAM06}
\defcitealias{Williams:2009}{WBC09}

\title[The Tully-Fisher relations of early-type discs]{The Tully-Fisher
relations of early-type spiral and S0 galaxies}

\author[Michael J. Williams et al.]{Michael J. Williams$^{1,2}$\thanks{E-mail:
williams@astro.ox.ac.uk}, Martin Bureau$^1$ and Michele Cappellari$^1$
\\$^1$Sub-Department of Astrophysics, University of Oxford, Denys Wilkinson Building, Keble Road, Oxford OX1 3RH
\\$^2$European Southern Observatory, Karl-Schwarzschild-Str. 2, D-85748
Garching bei M\"unchen, Germany}

\begin{document}

\date{Accepted 2010 July 22. Received 2010 July 21; in original form
2009 November 6}

\pagerange{\pageref{firstpage}--\pageref{lastpage}} \pubyear{2010}

\maketitle
\label{firstpage}

\begin{abstract} We demonstrate that the comparison of Tully-Fisher
relations (TFRs) derived from global \hi{} line widths to TFRs derived
from the circular velocity profiles of dynamical models (or stellar
kinematic observations corrected for asymmetric drift) is vulnerable to
systematic and uncertain biases introduced by the different measures of
rotation used. We therefore argue that to constrain the relative
locations of the TFRs of spiral and S0 galaxies, the same tracer and
measure must be used for both samples. Using detailed near-infrared
imaging and the circular velocities of axisymmetric Jeans models of 14
nearby edge-on Sa--Sb spirals and 14 nearby edge-on S0s drawn from a
range of environments, we find that S0s lie on a TFR with the same slope
as the spirals, but are on average $0.53\pm0.15$\,mag fainter at
$K_S$-band at a given rotational velocity. This is a significantly
smaller offset than that measured in earlier studies of the S0 TFR,
which we attribute to our elimination of the bias associated with using
different rotation measures and our use of earlier type spirals as a
reference. Since our measurement of the offset avoids systematic biases,
it should be preferred to previous estimates. A spiral stellar
population in which star formation is truncated would take $\approx
1$\,Gyr to fade by 0.53\,mag at $K_S$-band. If S0s are the products of a
simple truncation of star formation in spirals, then this finding is
difficult to reconcile with the observed evolution of the spiral/S0
fraction with redshift. Recent star formation could explain the observed
lack of fading in S0s, but the offset of the S0 TFR persists as a
function of both stellar and dynamical mass. We show that
the offset of the S0 TFR could therefore be explained by a systematic
difference between the total mass distributions of S0s and spirals, in
the sense that S0s need to be smaller or more concentrated than spirals.
\end{abstract}


\section{Introduction}
\label{sec:intro}

Both early- and late-type galaxies follow scaling relations. In the case
of elliptical galaxies, the radii, velocity dispersions and luminosities
lie on a Fundamental Plane \citep{Djorgovski:1987,Dressler:1987}. The
luminosities of disc galaxies scale with their sizes
\citep{Freeman:1970}, rotation speeds \citep{Tully:1977} and colours
\citep{Tully:1982}. The existence, scatter and evolution of these
scaling relations provides powerful constraints for models of galaxy
formation and evolution.

In this work we concentrate on the Tully-Fisher relation (TFR), first
discovered by \cite{Tully:1977}. In its original form, the TFR relates
the global \hi{} line widths of disc galaxies to their total
luminosities. The line width of a rotationally-supported galaxy is a
measure of the difference between the maximum rotation velocities of the
approaching and receding sides. A galaxy's luminosity is related to the
mass of its luminous component (and the evolutionary state of the
stellar populations). The observed correlation between global line
widths and luminosities is therefore a manifestation of a more general
connection between galaxy masses and rotational velocities. A
correlation is of course expected for gravitationally bound systems in
which the mass distribution is dominated by luminous matter.

Empirically, the correlation is tight enough to allow the TFR to be
inverted and used to determine distances \citep[e.g.][hereafter
TP00]{Sakai:2000,Tully:2000}, but systematic variations of the residuals
can arguably be used to infer the properties of dark haloes
\citep{Courteau:1999}. Efforts to simultaneously reproduce the
zero-point of the TFR and the galaxy luminosity function with
semi-analytical models of galaxy formation may reveal further
information on the role of the halo, the importance of feedback and the
evolution of the stellar populations
\cite[e.g.][]{Baugh:2006,Dutton:2007}. It is therefore clear that
investigating variations of the local TFR with galaxy luminosity, size,
type and other parameters should provide valuable insight into the
growth of luminous structures in the Universe.

There are systematic variations in the slope, intercept and scatter of
the TFR as a function of galaxy type. \cite{Roberts:1978} and
\cite{Rubin:1985} found that Sa spirals are fainter than Scs for a given
rotational velocity at optical wavelengths. At least some of this
difference has been attributed to variation in the shape of rotation
curves related to the bulge-to-disc ratio
\citep{Verheijen:2001,Noordermeer:2007}. The difference is much smaller
in the near-infrared \citep{Aaronson:1983,Peletier:1993}. This has been
taken as evidence that the discrepancy at optical wavelengths is at
least partially due to the effects of recent star formation on the
luminosity, which is reduced in the near-infrared. This effect has been
measured for a large, bias-corrected sample by \cite{Masters:2008}. They
find no evidence of any difference between the zero points of the Sb and
Sc TFRs and a small effect for Sa galaxies, which they find have a
variable offset compared to Sb\,--\,Sc spirals: $\approx 0.2$\,mag
brighter at $v \approx 150$\,\kms, no difference at $v \approx
180$\,\kms{} and $\approx 0.4$\,mag fainter at $v \approx 250$\,\kms.

Particular attention has been paid to S0 galaxies, which are the
earliest galaxies with rotationally-supported stellar discs. The fact
that they are more common in the centres of clusters, rarer on the
outskirts, and rarest of all in the field, the opposite of the trend
observed for spirals, suggests that environmental processes transform
spirals into S0s \citep{Spitzer:1951,Dressler:1980,Dressler:1997}.
Processes such as ram-pressure stripping \citep{Gunn:1972},
strangulation \citep{Larson:1980} and harassment \citep{Moore:1996} may
have removed the gas from the disks of spiral galaxies and left them
unable to form stars. S0s do however exist in the field, where such
processes should be insignificant. Passive evolution or the effects of
active galactic nuclei in spirals have been suggested as possible
formation mechanisms outside dense environments
\citep{van-den-Bergh:2009a}. Problems remain with the disk fading
picture, however: the bar fractions of spirals and S0s are discrepant
\citep{Aguerri:2009,Laurikainen:2009}, in a given environment S0s are at
least as bright as spirals \citep{Burstein:2005}, and the bulge-to-disc
ratios of S0s are difficult to reconcile with those of their presumed
progenitors \citep{Christlein:2004}. See \cite{Blanton:2009} for a
modern review of this issue.

Unless they are violent, these processes should make only a slight
change to the kinematics and dynamical masses of S0s, but a significant
change to their luminosities. Assuming they do not have systematically
different dark matter fractions, this change in stellar population
should cause S0s to have fainter luminosities for a given rotational
velocity and lie offset from the TFR of spirals. The size of any offset
and the difference between the scatters of the S0 and spiral TFRs can
therefore be used to constrain models of S0 formation and evolution
\citep[][hereafter BAM06]{Bedregal:2006}. A well-constrained S0 TFR also
raises the possibility of extending the Tully-Fisher distance
determination technique to earlier galaxy types, and of probing
variations of mass-to-light ratio with redshift.

Measuring the TFR of S0s is difficult because they do not usually have
extended gaseous discs. Instead of using a gaseous emission line width
as a measure of maximum rotation, several groups have used resolved
rotation curves derived from stellar absorption lines. These raw
observations must then be corrected for asymmetric drift (i.e. pressure
support) before they can be presumed to trace the circular velocity (and
therefore the potential of the galaxy). A number of groups have done
just this and found that S0s are indeed fainter than spirals for a given
rotational velocity. \cite{Neistein:1999} found that S0s lie on a TFR
offset by 0.5\,mag at $I$-band from the spiral TFR (with 0.7\,mag
intrinsic scatter). \cite{Hinz:2001} and \cite{Hinz:2003} found an
offset of 0.2\,mag at $I$ and $H$-band (with 0.5--1.0\,mag intrinsic
scatter). \citetalias{Bedregal:2006} combined these data with new observations of galaxies in
the Fornax cluster and found that the combined sample of 60 S0s lies
around 1.5\,mag below the spiral TFR at $B$-band (0.9\,mag scatter) and
1.2\,mag below the spiral TFR at $K_S$-band (1.0\,mag scatter). A
promising alternative to stellar absorption line kinematics is sparse
kinematic data from the emission lines of planetary nebulae. A pilot
study of the nearby S0 galaxy NGC~1023 with the Planetary Nebula
Spectrograph found that this particular S0 lies around 0.6\,mag below a
spiral TFR at $K_S$-band \citep{Noordermeer:2008}.

An alternative approach to measuring and correcting the rotational
velocities of the earliest type galaxies is to construct dynamical
models that, in principle, directly probe the circular velocity of a
system. At the expense of model-dependent assumptions, this approach
removes the uncertainties associated with the asymmetric drift
correction, which may be both significant and systematic.
\cite{Mathieu:2002} did this for a sample of six edge-on S0s and found
that they lie 1.8\,mag below an $I$-band spiral TFR (with 0.3\,mag
intrinsic scatter). Modelling also allows the TFR to be extended to
elliptical galaxies with little or no rotational support. As with S0s,
previous dynamical modelling studies of ellipticals have found that they
are fainter than spirals for a given rotational velocity:
\cite{Franx:1993}, 0.7\,mag at $R$-band; \cite{Magorrian:2001}, 0.8\,mag
at $I$-band; \cite{Gerhard:2001}, 0.6\,mag at $R$- and 1.0\,mag at
$B$-band; \cite{de-Rijcke:2007}, 1.5\,mag at $B$-, 1.2\,mag at
$K_S$-band.

However, all previous studies comparing the TFRs of S0s and ellipticals
to those of spirals have used independently determined reference TFRs
for spiral galaxies (taken from, e.g., \citealt{Sakai:2000}; TP00).
These reference TFRs are derived in the traditional way, using global
line widths from \hi{} observations. In \refsec{sec:rotation} we discuss
the technical and observational issues involved in characterizing the
enclosed mass of a disc galaxy in an unbiased way. We argue that
comparisons between TFRs derived from different measures of rotation are
not a priori justified and are likely to introduce artificial and
uncertain offsets in practice. 

The main astrophysical goal of the present work is thus to compare the
TFRs of spiral and S0 galaxies in a way that is not vulnerable to these
uncertain systematics, by using measures of luminosities and rotational
velocities determined identically for both S0s and spirals. Readers who
are more interested in the results of this comparison and its
implication for the evolution of S0s, may wish to skip ahead to
\refsec{sec:data}, where we present our sample. In \refsec{sec:fitting}
we describe the numerical method used to simultaneously determine the
TFRs of spirals and S0s. We then present TFRs for the two samples in
terms of near-infrared and optical luminosities, stellar mass and
dynamical mass. In \refsec{sec:discussion} we discuss our results in the
context of models of S0 formation and evolution and previous analyses of
S0 TFRs. We conclude briefly in \refsec{sec:conclusion}.

\section{Measures of rotation} \label{sec:rotation}

\subsection{The problem}
\label{sec:problem}

To construct a Tully-Fisher relation one needs a measure of rotation.
This can be derived from, for example, the global \hi{} line width,
spatially-resolved \hi{}, \halpha{} or stellar rotation curves or
velocity fields, sparse tracers (e.g. globular clusters and planetary
nebulae), or the circular velocity of dynamical models at some fiducial
location. In principle, all these measures are related to the rotation
of the galaxy, but the relationship of these single numbers to the
enclosed mass is complicated by the fact that observed rotation curves
(and indeed modelled circular velocity curves) do not always flatten.
Moreover, it not obvious that these observational measures directly and
accurately comparable to each other. Both the nature of the observations
and the corrections applied for different techniques could conceivably
introduce systematics that vary with galaxy properties such as type,
mass or size. 

For example, the transformation of a global \hi{} line width to a
measure of rotation involves corrections for instrumental broadening and
turbulent motion \citep[e.g.][]{Tully:1985,Verheijen:2001}. The extent
to which these corrections are able to recover the intrinsic motion of
the tracer or the true circular velocity of the galaxy potential has
recently been questioned \citep{Singhal:2008}. Meanwhile, stellar
kinematics must be corrected for asymmetric drift, which typically
involves making approximations which are difficult to justify prima
facie (see \refsec{sec:defined}). Finally, the construction of dynamical
models often involves simplified modelling of the full distribution
function \citep[e.g.][]{Mathieu:2002} or assumptions about intrinsic
morphology, anisotropy and the role of dark matter \cite[e.g.][hereafter
WBC09]{Williams:2009}. 

These considerations place the direct comparisons of TFRs derived using
different measures of rotation on shaky ground. Offsets which are
ascribed to fading or brightening could in fact be due to systematic
biases that differ between the measures of rotation used. For typical
slopes of the TFR, a relatively small systematic difference in velocity
of $\sim 0.1$\,dex introduces an offset in the TFR that is
indistinguishable from a $\sim 1$\,mag difference in luminosity. This is
of the order of the typical zero-point offsets claimed for S0 TFRs
\citep{Neistein:1999,Hinz:2001,Hinz:2003,Mathieu:2002,Bedregal:2006}, so
it is crucial that we eliminate (or at least quantify) the systematics
introduced by such comparisons. In the remainder of this section, we
examine the differences between the results of three different measures
of rotation when applied to the same 28 early-type disc galaxies, and
discuss the implications for comparisons of the zero-point of the TFR.

\subsection{Four possible measures of rotation defined}
\label{sec:defined}

We aim to compare measures of rotation derived from global
\hi{} line widths (which we denote \vhi), resolved gas rotation curves
(\vgas), stellar kinematics (\vdrift)
and dynamical models ($\vmodel$) for the 28 galaxies in the present work.
This sample, which consists of 14 spirals and 14 S0s, is described in
\refsec{sec:data}. 

For \vhi, the rotational velocity derived from global \hi{} line widths,
we adopt the quantity \texttt{vrot} from HYPERLEDA \citep{Paturel:2003}.
This is the inclination-corrected maximum gas rotation velocity, based
on an average of independent \hi{} line width determinations taken from
the literature. Strictly speaking, HYPERLEDA also uses
spatially-resolved \halpha{} rotation curves to calculate \texttt{vrot},
but these observations were only present in HYPERLEDA for four of the
galaxies in our sample, and there was no evidence that they
systematically disagreed with the \hi{} line width values at any more
than the 0.01\,dex level.

For \vgas, we fit Gaussians to the position-velocity diagrams (PVDs)
presented in \cite{Bureau:1999}. These are derived from [N\,\textsc{ii}]
emission lines. We only use the region of the PVD where the rotation
curve is flat. This restriction means that it is not possible to measure
\vgas{} for many of the galaxies in our sample, either because emission
was not detected, or because it was not sufficiently extended.

\begin{figure*}
\includegraphics[width=17.5cm]{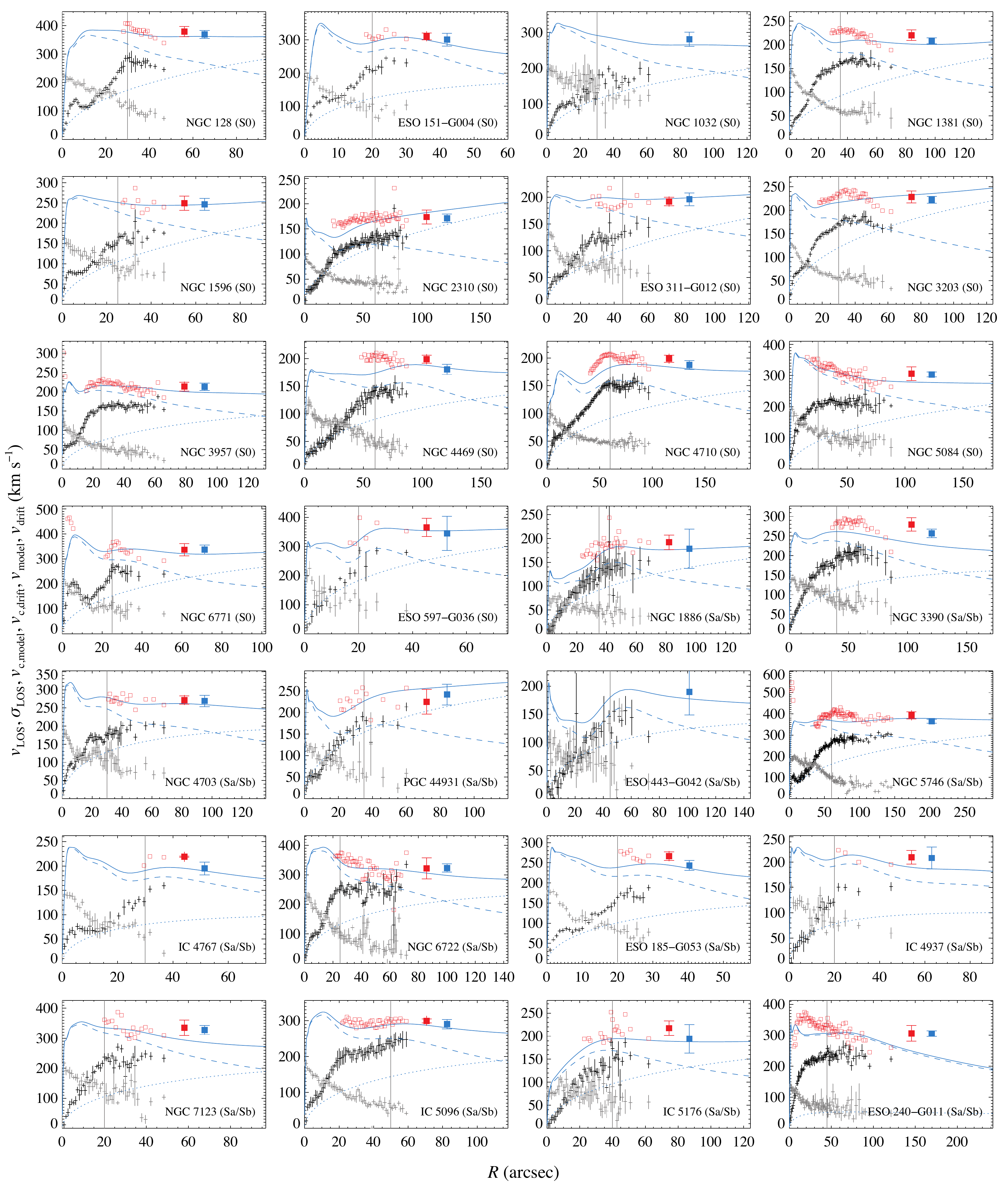}
\caption{Observed, corrected and modelled kinematic data for all sample
galaxies. The black points are the observed mean line-of-sight stellar
velocities $v_\mathrm{LOS}$ and the grey points the line-of-sight
velocity dispersions $\sigma_\mathrm{LOS}$, both from
\protect\cite{Chung:2004}. The open red squares are estimates of the
circular velocity derived by correcting the observations for
line-of-sight and asymmetric drift effects ($v_\mathrm{c,drift}$). There
are missing points at small radii because the correction is unreliable
where the local dispersion is large compared to the line-of-sight
velocity. The solid blue line is the circular velocity of the galaxy
dynamical model from \citetalias{Williams:2009} ($v_\mathrm{c,model}$).
These models consist of contributions from a stellar component (dashed
blue line) and a dark halo (dotted blue line). The vertical line is the
radii beyond which the observed rotation curve is flat, which is
important for our calculation of \vdrift{} and \vmodel{}, the single
values used to characterize rotation. These are determined as described
in the text. Their values for each galaxy are shown as filled red and
blue squares respectively, at arbitrary large radii.}
\label{fig:models}.
\end{figure*}

We calculate the asymmetric drift-corrected stellar velocity, \vdrift{},
using the observed stellar kinematics presented in \cite{Chung:2004}. In
order to more accurately measure the local circular velocity we attempt
to correct these observations for the effects of line-of-sight
integration and asymmetric drift using the same recipe as used in
\citetalias{Bedregal:2006}, which we describe briefly below. We make no
attempt to improve this recipe, since one of our goals is to assess the
biases in previous determinations of the TFR based on stellar
kinematics. The raw observations and their corrected values are shown in
\reffig{fig:models}.

The line-of-sight correction for an edge-on disc assumes that the local
azimuthal velocity $v_\phi(R)$ at a galactocentric distance $R$ (in the
galaxy's cylindrical coordinate system $(R, \phi, z)$) is related to
$v_\mathrm{LOS}(x')$, the luminosity-weighted mean line-of-sight velocity at a
projected distance from the galaxy centre $x'$, by the equation  
\begin{equation}
\label{eqn:los}
v_\mathrm{LOS}(x') = \frac{\int_0^\infty v_\phi(R) \rho(R) \cos\phi\,\ud
z'}{\int_0^\infty \rho(R)\,\ud z'}.
\end{equation}
Following the notation of, e.g., \cite{Binney:2008}, $(x', y', z')$ is
the observer's coordinate system on the sky, in which the $x'$ axis is
aligned with the projected major axis of the galaxy and the $z'$ axis is
along the line of sight. $\rho(R)$ is the luminosity density. The
integral is evaluated along that line-of-sight $z'$. By assuming that
the local streaming velocity $v_\phi$ is independent of $R$, one can use
the above relationship to infer it. This assumption is valid only if $R$
is at least the radius at which the intrinsic circular velocity curve flattens.
To evaluate these integrals we assume $\rho(R)$ is an exponential disc
of scale length $R_\mathrm{d}$, which we measure from the major-axis
surface brightness profiles of our $K_S$-band images at radii outside
the bulge \citep{Bureau:2006}. 

We then apply a correction for asymmetric drift to $v_\phi(R)$ to
derive $v_\mathrm{c,drift}(R)$, an estimate of the true local circular velocity:
\begin{equation}
\label{eqn:drift}
v_\mathrm{c,drift}^2(R) = v_\phi^2(R) + \sigma_\phi^2(R)\left(2\frac{R}{R_\mathrm{d}} -
1\right),
\end{equation}
where $\sigma_\phi(R)$ is the azimuthal stellar velocity dispersion.
This equation is derived from the Jeans equations (\citealt{Jeans:1922};
\citealt{Binney:2008}) by assuming a thin disc in a steady state and the
epicyclic approximation ($\sigma_\phi^2/\sigma_R^2 = 0.5$ for
constant $v_\phi$). To estimate $\sigma_\phi(R)$, we assume that it is
equal to an exponential fit to the observed stellar line-of-sight
velocity dispersion $\sigma_\mathrm{LOS}(x')$. This assumption is of course
flawed, since the observed dispersion includes a contribution from the
changing component of the azimuthal velocity along the line-of-sight.
This effect will bias the asymmetric drift correction to be too high.
However, the error is small in practice, and, crucially for the present
work, it is not a strong function of galaxy type in the range S0--Sb.
Under the assumptions described above, for disks with typical levels of
pressure support ($v_\phi/\sigma_\phi \sim 5$ at the relevant radii
$\approx 2$\,--\,$4\,R_\mathrm{d}$), numerical integration shows that
the error introduced by wrongly assuming $\sigma_\phi(R) =
\sigma_\mathrm{LOS}(x')$ causes $v_\mathrm{c,drift}$ to overestimate the
true circular velocity by $\approx 0.04\pm0.01$\,dex. This error is
almost independent of $v_\phi/\sigma_\phi$ in the range $2 \le
v_\phi/\sigma_\phi \le 20$, a quantity which is in any case independent
of morphological type for our sample.

Finally, the local circular velocity of the mass model,
$v_\mathrm{c,model}$, is derived from the dynamical models presented in
\citetalias{Williams:2009}. In that work, we modelled the mass
distribution of each galaxy by assuming it is composed of an
axisymmetric stellar component with a constant mass-to-light ratio and a
spherical NFW dark halo \citep{Navarro:1997}. The axisymmetric stellar
component is based on an analytic fit to the observed near-infrared
photometry of the composite bulge and disk. The absence of a triaxial
bar component in our models does not significantly affect either the
global properties of the models or the local kinematics at the large
radii relevant to the TFR (see section 5.5.3 of
\citetalias{Williams:2009} for a more extensive discussion of this
issue). We assumed a particular relationship between halo mass and
concentration \citep{Maccio:2008}. We determined the three parameters of
the models (the stellar mass-to-light ratio, virial halo mass and
orbital anisotropy) by comparing the observed stellar second velocity
moment $v_\mathrm{rms}(x') = [v_\mathrm{LOS}^2(x') +
\sigma_\mathrm{LOS}^2(x')]^{1/2}$ along the major axis to that predicted
by solving the Jeans equations assuming a constant anisotropy
(parametrized as $\beta_z = 1 - \sigma_z^2/\sigma_R^2$). The solution of
the Jeans equations under these assumptions is described in
\cite{Cappellari:2008}. Note that while $v_\mathrm{c,drift}$ described
above is weakly affected by our assumptions that
$\sigma_\phi^2/\sigma_R^2 = 0.5$ and $\sigma_\phi(R) =
\sigma_\mathrm{LOS}(x')$, $v_\mathrm{c,model}$ suffers from no such
biases. Our solutions of the Jeans equation are constrained by a fit to
the observed second velocity moment along the line-of-sight. The second
moment is strictly independent of $\sigma_\phi^2/\sigma_R^2$ so no
assumption for this ratio is required, and the implementation rigorously
accounts for line-of-sight integration.

To construct a TFR, the radial profiles of the asymmetric-drift
corrected stellar kinematics, $v_\mathrm{c,drift}$ and of the model
circular velocity, $v_\mathrm{c,model}$ must be characterized by a
single number, \vdrift\ and \vmodel\ respectively. We do this by
determining the flat region of the observed rotation curve by eye. We
then take \vdrift{} to be the mean of the values of $v_\mathrm{c,drift}$
beyond this point. The continuous Jeans modelled circular velocity curve
is evaluated at the same radii as the observed stellar kinematics. Its
mean in the flat region of the observed rotation curve gives \vmodel\.
Of course, it is possible to evaluate the circular velocity of the
models at arbitrary radii, but by restricting ourselves to those radii
for which we have stellar kinematic observations, we avoid using the
mass models in regions where they are not constrained by the data.
\reffig{fig:models} shows these calculations for all sample galaxies.

\subsection{Comparison}
\label{sec:comparison}

\begin{figure}
\includegraphics[width=8.4cm]{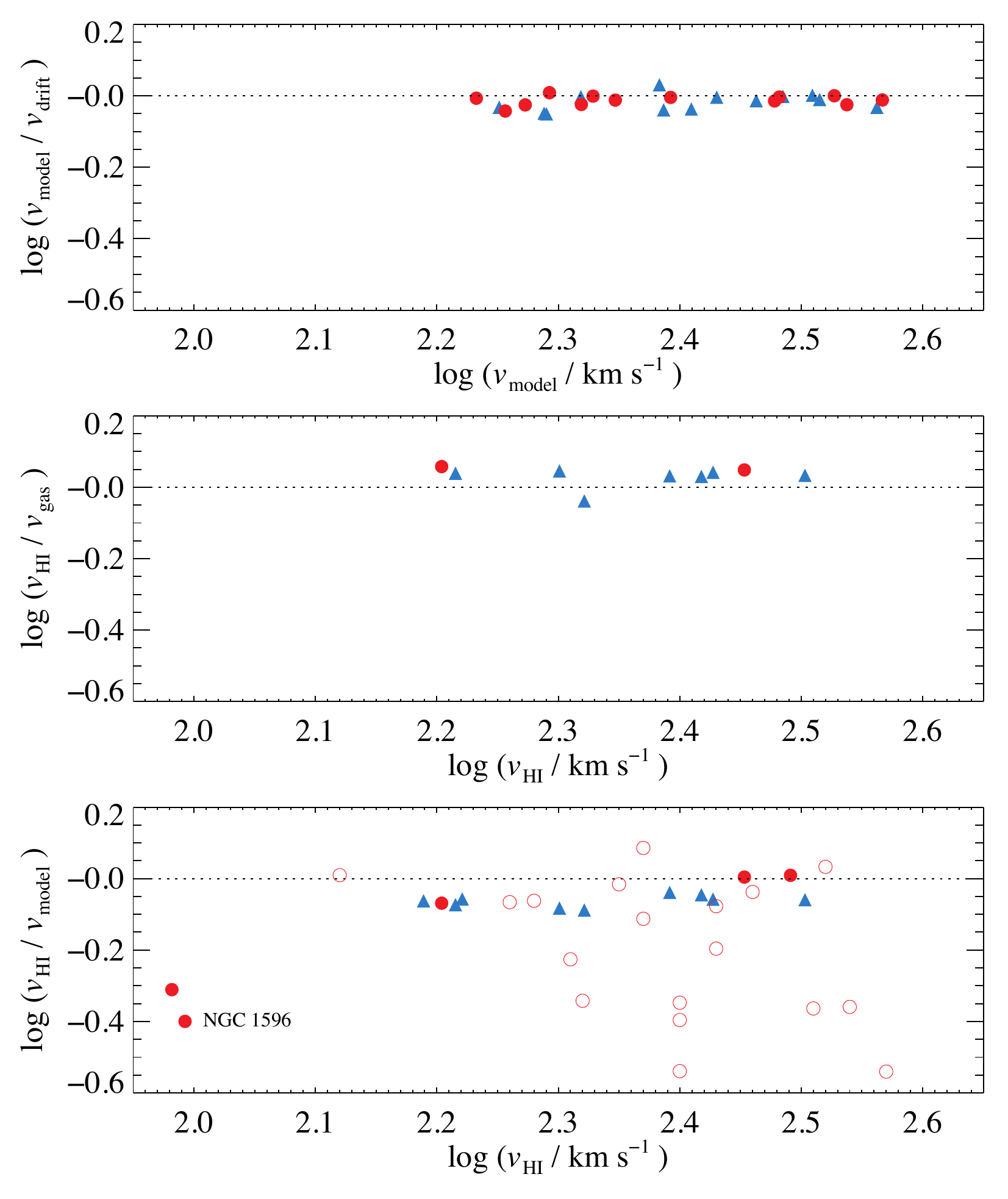}
\caption{Comparisons of the circular velocities of the mass models
(\vmodel), observed stellar kinematics corrected for asymmetric
drift (\vdrift) and rotational velocities derived from global \hi{} line
widths (\vhi) and ionized gas position-velocity diagrams (\vgas). The
solid blue triangles are the Sa--Sb galaxies from our sample and the
solid red circles the S0s. The open red circles are
the S0s presented in \citetalias{Bedregal:2006}. They do not estimate \vmodel{} so, in order
to place these galaxies on the plot, we assume $\vdrift \approx \vmodel$
for them.} 
\label{fig:vall}
\end{figure}

A comparison of the four measures of rotation is shown in
\reffig{fig:vall}. The upper panel demonstrates that
$\left<\log(\vmodel/\vdrift)\right>$ $= -0.015\pm0.003$\,dex with an rms scatter
of 0.019\,dex for both spirals and S0s. This small difference, which is
on average less than 4 per cent, is probably due to the additional
approximations and assumptions introduced in the asymmetric drift
correction compared to the Jeans models. We note in particular that it
is consistent with the error due to assuming $\sigma_\phi(x') =
\sigma_\mathrm{LOS}(x')$, as discussed in \refsec{sec:defined}. The
absolute offset between these measures is not significant in the context
of the TFR and, crucially, there is no evidence of any systematic
difference between S0s and spirals. The scatter is also extremely small.
Since both \vdrift{} and \vmodel{} are derived from the observed stellar
kinematics, their close agreement is not of itself independent proof of
the accuracy with which they trace the true circular velocity. The
agreement of \vdrift{} with the results of detailed and robust Jeans
modelling does suggest, however, that the approximations inherent in
eqns.~(1--2) yield an approximately valid solution of the Jeans
equations.

The middle panel demonstrates fairly good agreement between the two
measures based on gas, \vhi{} and \vgas{}.
$\left<\log(\vhi/\vgas)\right> = 0.032\pm0.009$\,dex with a rms scatter
of 0.029\,dex. There is some evidence that \vhi\ systematically exceeds
\vgas\ by 5\,--\,10 per cent. Statistics are poorer because gas is rare
in S0s by definition, but there is no evidence of any systematic
difference between $\vhi/\vgas$ in S0s and spirals.

However, the bottom panel of \reffig{fig:vall} shows that $\vmodel$ (and
therefore \vdrift) does not in general agree \vhi{} (and thus \vgas):
$\left<\log(\vhi/\vmodel)\right> = -0.094\pm0.031$\,dex with a rms
scatter of 0.115\,dex (or $-0.18\pm0.043$\,dex with a rms scatter of
0.243\,dex if the \citetalias{Bedregal:2006} galaxies are included).
Some of this difference and scatter is caused by pathological cases of
\vhi\ derived from single dish measurements, i.e. spatially unresolved
measurements of 21\,cm signal from galaxies with a warped gas disc,
polar rings or gas-rich nearby companions that fall within the telescope
beam (e.g. this is almost certainly the case for the S0 galaxy NGC~1596,
see \citealt{Chung:2006}).

But even when possible pathological cases are avoided, for example by
ignoring S0 galaxies, which are more likely to be problematic, there
remains a clear systematic offset: $\left<\log(\vhi/\vmodel)\right> =
-0.063\pm0.005$\,dex (15 per cent) with 0.02\,dex rms scatter for the
spirals in our sample. For these galaxies, the rotational velocity
derived from the global \hi{} line width is significantly and
systematically smaller than that derived from stellar kinematics or
dynamical modelling. Because \vhi\ is derived from spatially unresolved
data, this could be explained by the global \hi\ line width probing the
local gas velocity out to different radii than our typical stellar
kinematic data \citep{Noordermeer:2007a}.

\begin{figure}
\includegraphics[width=8.4cm]{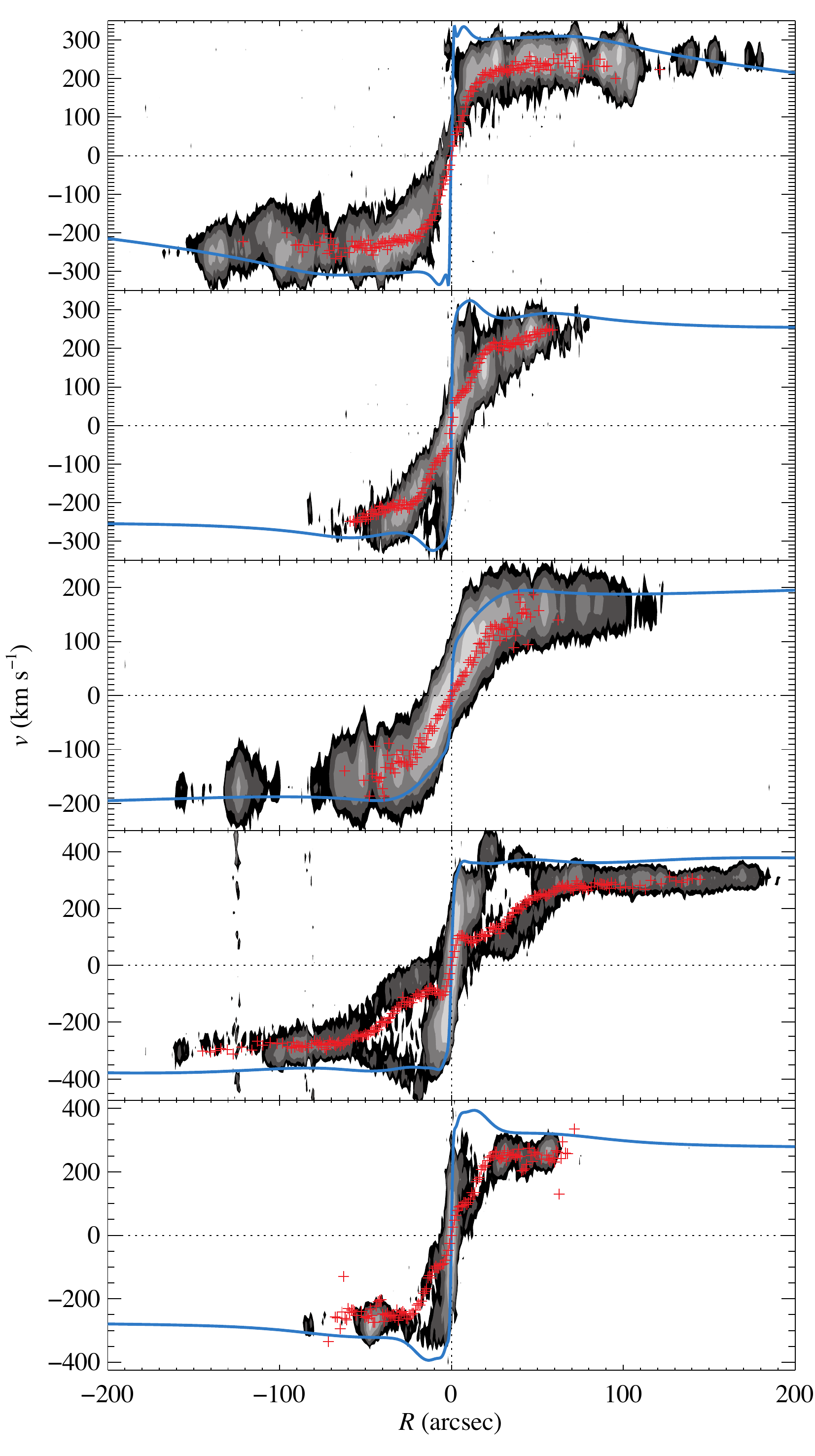}
\caption{Comparisons of [N\,\textsc{ii}] $\lambda6584$ position--velocity
diagrams, stellar kinematics, and the circular velocity implied by Jeans
modelling for the five galaxies in the sample with optical emission that
extends to the disc. From top to bottom: ESO~240-G11, IC~5096, IC~5176,
NGC~5746 and NGC~6722. The filled contours are the continuum-subtracted
optical spectra originally presented in \protect\cite{Bureau:1999}. The
observed mean stellar velocities \protect\citep{Chung:2004} and the circular
velocity profiles inferred from detailed Jeans modelling of the light
distribution and stellar kinematics \citepalias{Williams:2009} are overplotted as red
crosses and blue lines respectively.}
\label{fig:gaspvds}
\end{figure}

We find a similar systematic difference between \vmodel\ (or \vdrift)
and \vgas: $\left< \log(\vgas/\vmodel) \right> = -0.086 \pm 0.008$.
\vgas\ is derived from a spatially resolved PVD taken from the same
long-slit optical spectrum with which we derive the stellar kinematics.
We can therefore verify directly that it does not suffer from either
pathological problems related to gas-rich companions, warps or polar
rings, or doubt about the radii probed compared to our stellar
kinematics. We show this in \reffig{fig:gaspvds} for the five sample
galaxies with optical emission that extends to the disk. It is clear
that the stars and gas are associated with the same galaxy and are
observed at similar radii. Therefore, unless light-emitting gas is
absent from the tangent point along the line-of-sight in these edge-on
galaxies (a possibility we cannot rule out), the emission line gas
kinematics and absorption line stellar kinematics should imply the same
local circular velocity and the same single characteristic measure of
rotation, (\vgas, \vdrift\ or \vmodel).

But in these five galaxies galaxies, the stars and gas rotate with
essentially the same velocity as a function of position in the outer,
flat parts of the observed rotation curve. The stars have non-negligible
dispersion so we know they do not trace the local circular velocity by
an amount that can be calculated using the Jeans equations. Given this
observation, it is not surprising that \vgas\ is significantly less than
\vmodel. 

In summary, while agreement between \vgas\ and \vhi\ is fairly good, and
from this we conclude that \vmodel{} and \vdrift{} (based on stars seen
in absorption) do not imply the same circular velocity as \vhi{} and
\vgas{} (based on gas seen in emission). This is not a new result. In
early-type disc galaxies, the observed stellar kinematics and the
well-understood Jeans equations often imply a local circular velocity
greater than the observed rotational velocity of gas
\citep[e.g.][]{Kent:1988,Kormendy:1989,Bertola:1995,Cinzano:1999,Corsini:1999,Cretton:2000,Vega-Beltran:2001,Pizzella:2004,Krajnovic:2005,Weijmans:2008,Young:2008}.
This puzzle is made more acute by the small velocity dispersions
observed in the emission line PVDs. We will return to the gas dispersion
in a future work.

A definitive resolution of this issue is outside the scope of the
present work and, fortunately, it is unnecessary for our goals. For the
purposes of comparing TFRs (or for determining relative distances; see
e.g. \citealt{Courtois:2009}), it does not matter which measure is
`right', as long as we use the same measure for the two samples and the
measure chosen does not introduce a systematic bias between samples. We
see no evidence of this in our data, for any of the four measures
explored.

Our conclusions in this section have three consequences for previous
works that measure the TFR of S0 galaxies using stellar kinematics
corrected for asymmetric drift
\citep[e.g.][\citetalias{Bedregal:2006}]{Neistein:1999,Hinz:2001,Hinz:2003}:
(i) the asymmetric drift correction they use yields similar results to
detailed Jeans modelling; (ii) the drift correction does not seem to
introduce a systematic bias between spirals and S0s and, if applied to
samples of both spirals and S0s, can indeed be used to compare the TFRs
of the two classes; (iii) however, a TFR derived from asymmetric
drift-corrected stellar kinematics can\emph{not} be directly compared to
TFRs derived from global \hi{} line widths or resolved emission line
PVDs, as these previous authors did. As we argue above, such comparisons
are not a priori justified and, in the case of both our sample and the
larger sample of \citetalias{Bedregal:2006}, they appear to introduce
systematic errors that, on a Tully-Fisher plot, are of the same order
(and in the same direction) as the TFR offsets that are usually
interpreted as luminosity evolution. The conclusions of previous works
regarding the offset of the S0 TFR should therefore be treated with
caution. In addition, given the large scatter we observe in
$\log(\vhi/\vmodel)$ and $\log(\vhi/\vdrift)$, it is far from clear that
any simple additive or multiplicative correction could eliminate the
bias.

For our sample, we are able to side-step this issue completely by using
a single measure of rotation for both spirals and S0s. In this way we
compare the TFRs of S0, Sa and Sb galaxies in a manner which is not
subject to the possible systematics discussed above. We adopt \vmodel{}
as our fiducial measure of rotation for \emph{both} spirals \emph{and}
S0s. The figures and discussion that follow use only this measure. 

\section{Sample and data}
\label{sec:data}

\begin{table*}
\caption{Sample.}
\label{tab:sample}
\begin{tabular*}{\textwidth}{@{\extracolsep{\fill}}lccccccccccl}
\hline
\noalign{\smallskip}
Galaxy & $M_{K_S}$ & $M_B$ & $\log(M_\star/M_\odot)$ & $\log(M_\mathrm{dyn}/M_\odot)$ & $R_\mathrm{d}$ & $\vmodel$ & $\vdrift$
& $\vhi$ & $\vgas$ & 2MRS LDC & Cluster \\
       &  (mag) & (mag) & & & (arcsec) & (km\,s$^{-1}$) & (km\,s$^{-1}$) & (km\,s$^{-1}$) & (km\,s$^{-1}$) &  group size \\
(1)    &   (2) & (3) & (4) & (5) & (6) & (7) & (8) & (9) & (10) & (11) & (12) \\
\noalign{\smallskip}
\hline
\noalign{\smallskip}
\multicolumn{12}{c}{S0} \\
\noalign{\smallskip}
     NGC 128 & $-25.35$ & $-21.40$ &  11.47 &  11.56 &   16.7 &    369 &    379 & \ldots & \ldots & \ldots \\
ESO 151-G004 & $-24.85$ & $-20.41$ &  11.40 &  11.46 &    8.2 &    301 &    311 & \ldots & \ldots & $^a$ \\
    NGC 1032 & $-24.47$ & $-20.71$ &  11.13 &  11.22 &   33.0 &    281 & \ldots &    284 &    254 & \ldots \\
    NGC 1381 & $-22.76$ & $-18.77$ &  10.49 &  10.59 &   12.3 &    208 &    220 & \ldots & \ldots & 43 & Fornax \\
    NGC 1596 & $-22.86$ & $-18.98$ &  10.56 &  10.66 &   20.5 &    247 &    249 &     98 & \ldots & 30 \\
    NGC 2310 & $-22.48$ & $-19.43$ &  10.25 &  10.48 &   25.2 &    171 &    173 & \ldots & \ldots & \ldots \\
ESO 311-G012 & $-23.06$ & $-19.92$ &  10.45 &  10.53 &   28.3 &    196 &    192 &     96 & \ldots & \ldots \\
    NGC 3203 & $-23.89$ & $-19.99$ &  10.76 &  10.94 &   16.3 &    222 &    228 & \ldots & \ldots & 210  & Hydra \\
    NGC 3957 & $-23.29$ & $-19.20$ &  10.76 &  10.85 &   19.1 &    213 &    213 & \ldots &    178 & 13 \\
    NGC 4469 & $-23.01$ & $-19.02$ &  10.56 &  10.68 &   16.5 &    180 &    199 & \ldots & \ldots & 300 & Virgo \\
    NGC 4710 & $-23.47$ & $-19.40$ &  10.56 &  10.68 &   23.6 &    187 &    199 &    160 &    140 & 300 & Virgo \\
    NGC 5084 & $-24.73$ & $-21.85$ &  11.16 &  11.22 &   34.7 &    303 &    306 &    310 & \ldots & 15 \\
    NGC 6771 & $-25.08$ & $-20.81$ &  11.40 &  11.46 &   16.2 &    337 &    337 & \ldots &    284 & 97 \\
ESO 597-G036 & $-25.36$ & $-20.70$ &  11.58 &  11.73 &   14.4 &    345 &    365 & \ldots & \ldots & 3 \\
\noalign{\smallskip}
\multicolumn{12}{c}{Sa--Sb} \\
\noalign{\smallskip}
    NGC 1886 & $-22.90$ & $-19.73$ &  10.47 &  10.63 &   14.5 &    178 &    192 &    155 & \ldots & \ldots \\
    NGC 3390 & $-24.90$ & $-21.49$ &  11.19 &  11.27 &   20.4 &    257 &    279 &    210 &    229 & 210 & Hydra \\
    NGC 4703 & $-25.43$ & $-20.92$ &  11.38 &  11.49 &   29.2 &    269 &    272 &    246 &    229 & 80 & \\
   PGC 44931 & $-24.46$ & $-21.33$ &  11.07 &  11.27 &   21.1 &    242 &    225 &    200 &    180 & 80 & \\
ESO 443-G042 & $-23.49$ & $-20.21$ &  10.76 &  10.90 &   11.4 &    190 & \ldots &    166 & \ldots & 38 \\
    NGC 5746 & $-25.64$ & $-22.30$ &  11.63 &  11.75 &   69.1 &    365 &    393 &    319 &    295 & 45 \\
     IC 4767 & $-23.69$ & $-19.72$ &  10.93 &  10.97 &    8.2 &    195 &    219 & \ldots & \ldots & 97 \\
    NGC 6722 & $-25.38$ & $-21.97$ &  11.49 &  11.60 &   12.5 &    323 &    322 & \ldots &    263 & 97 \\
ESO 185-G053 & $-24.22$ & $-20.09$ &  11.01 &  11.06 &    5.0 &    244 &    267 & \ldots & \ldots & 42 \\
     IC 4937 & $-24.67$ & $-20.46$ &  11.18 &  11.24 &   18.7 &    208 &    210 & \ldots & \ldots & 3 \\
    NGC 7123 & $-25.29$ & $-20.81$ &  11.35 &  11.40 &   19.2 &    327 &    335 & \ldots & \ldots & \ldots \\
     IC 5096 & $-24.93$ & $-21.21$ &  11.27 &  11.36 &   14.5 &    290 &    300 &    262 &    244 & 5 \\
     IC 5176 & $-23.31$ & $-19.95$ &  10.60 &  10.72 &   20.9 &    194 &    217 &    164 &    150 & \ldots \\
ESO 240-G011 & $-24.79$ & $-21.56$ &  11.56 &  11.57 &   35.7 &    305 &    306 &    268 &    243 & 3 \\
\noalign{\smallskip}
\hline
\noalign{\smallskip}
\end{tabular*}
\begin{minipage}{\textwidth}
\emph{Notes.} Sample divided into S0s and spirals based on
classifications of \cite{Jarvis:1986}, \cite{de-Souza:1987},
\cite{Shaw:1987} and \cite{Karachentsev:1993}.
(1) Galaxy name. (2)--(3) Total $K_S$-band and $B$-band absolute
magnitude. (4)--(5) Total stellar and dynamical mass. (6) Disk scale
length measured using $K_S$-band images presented in \cite{Bureau:2006}.
(7) Circular velocity derived from the best-fit mass model presented in
\citetalias{Williams:2009}. (8) Rotational velocity derived from the observed stellar
kinematics presented in \cite{Chung:2004}, corrected for line-of-sight
effects and asymmetric drift. Missing values are cases where the drift
correction was too large to be considered reliable. (9) Rotational
velocity derived from the global \hi{} line width, i.e. the quantity
\texttt{vrot} from HYPERLEDA where available. (10) Ionized gas
rotational velocity measured by a Gaussian fit to [N\,\textsc{ii}]
position-velocity diagrams presented in \cite{Bureau:1999}. (11) Number
of companion group members in the Low Density Contrast 2MRS Group
Catalog \citep{Crook:2007}. Ellipses indicate
that a galaxy was not found in the catalog, implying that it had 2 or
fewer nearby companions. ($^a$) ESO~151-G004 is not present in the 2MASS
Extended Source Catalog \citep{Jarrett:2000} and so is not included in
the 2MRS database. (12) Cluster membership. For
full details of how the quantities in Columns (2)--(10) were determined,
see \refsec{sec:defined} and \refsec{sec:data}.
\end{minipage}
\end{table*}

To make the reliable comparison we described in the previous section, we
use the sample of 14 early-type spirals (Sa and Sb) and 14 S0s presented
in Table~\ref{tab:sample}. Dynamical models of these galaxies were
constructed in \citetalias{Williams:2009}. All of the galaxies are oriented
close to edge-on (within $\approx 5^\circ$). Many of the galaxies have
boxy bulges, which are believed to be bars viewed side-on
\cite[e.g.][]{Kuijken:1995,Merrifield:1999,Bureau:1999,Chung:2004}, and
it was to probe these bulges that the sample was constructed and
observed. The bars should not, however, affect the results we present
here because we use measures of rotational velocity well outside the
boxy bulge region. The fraction of galaxies with boxy bulges in the sample
($\approx 75$ per cent) is not in any case significantly different to
the fraction of barred galaxies in the local universe \citep[$\approx$
65 per cent; see, e.g.][]{Eskridge:2000,Whyte:2002,Marinova:2007}.

The sample was selected without regard for environment, and so includes
galaxies in a wide range of environments. Two S0s are members of the
Virgo cluster, an S0 and a spiral are members of the Hydra cluster, and
an S0 is a member of the Fornax cluster. Cross-referencing our sample
with the 2MASS Redshift Survey (2MRS) Group Catalog \citep{Crook:2007}
demonstrates that the remainder of the sample is made up of members of
intermediate-size groups and relatively isolated galaxies.

All the galaxies are relatively bright and fast-rotating, and the
spirals and S0s cover the same range of luminosities. This means that we
are not vulnerable to variation in the slope of the TFR at the low or
high mass end, as has been observed
\citep[e.g.][]{Peletier:1993,Noordermeer:2007}. We further note that the
sample contains some very bright S0s, such as ESO~151-G004 and NGC~6771,
that have large boxy bulges. If these bulges are in fact bars, and
therefore the products of secular evolution, then these particular
bright S0s are unlikely to be the elliptical-like products of major
mergers, as has previously been suggested
\citep{Poggianti:2001,Mehlert:2003,Bedregal:2008,Barway:2009}.

\begin{figure}
\includegraphics[width=8.4cm]{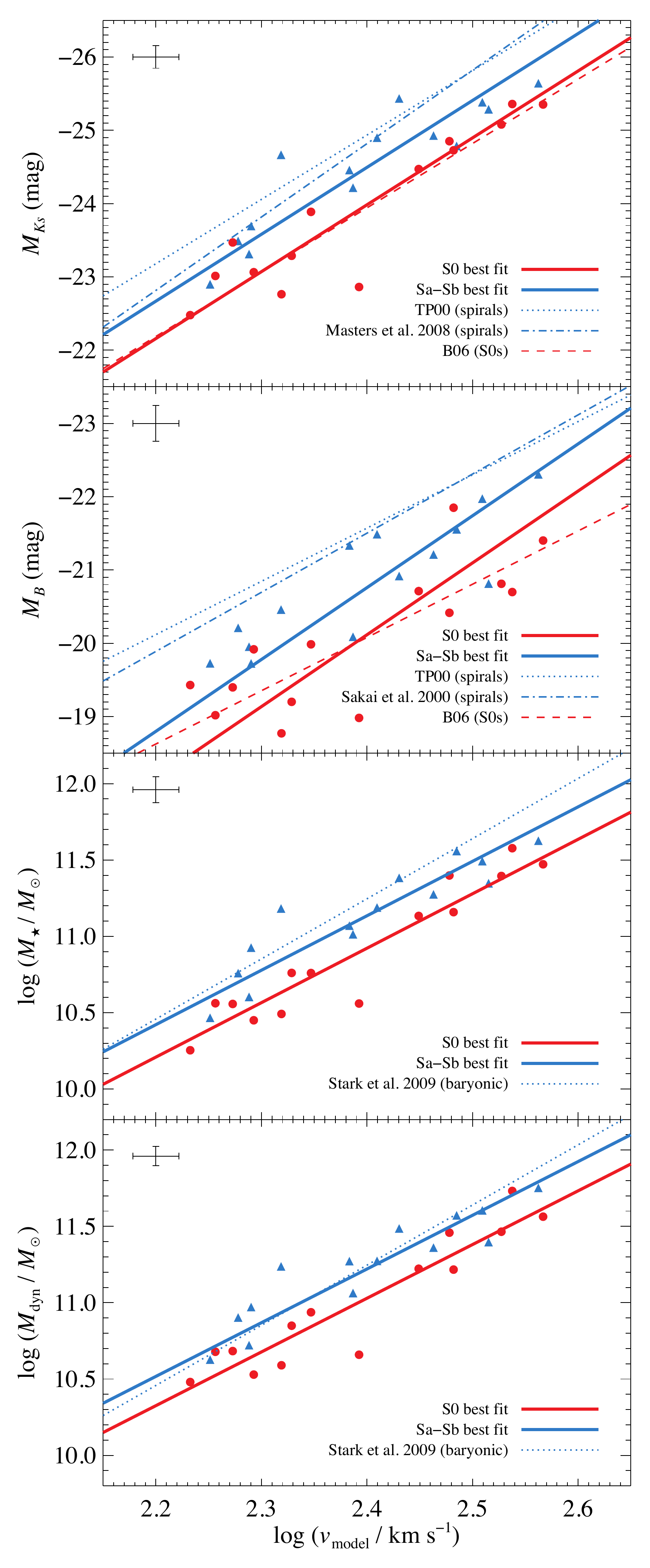}
\caption{Tully-Fisher relations for our samples of S0 and spiral
galaxies. From top to bottom, as a function of \vmodel{}: total absolute
$K_S$-band luminosity, total absolute $B$-band luminosity, total stellar
mass and total dynamical mass. Spiral galaxies are shown as blue
triangles and S0
galaxies as red circles. A median error bar is shown in the top left corner
of each plot. The best fit relations found by inverse regression are
shown as solid lines (blue for spirals, red for S0s). Their parameters
are given in Table~\ref{tab:params}. The dashed, dotted and dot-dashed
lines are a selection of TFRs found by previous authors for different
morphological types. See text for a full discussion of these relations.}
\label{fig:tf}
\end{figure}

In \reffig{fig:tf} we present the spiral and S0 TFRs as functions of
$K_S$-band luminosity, $B$-band luminosity, stellar mass and dynamical
mass, all discussed below. Throughout this work we assume an absolute
solar magnitude $M_{Ks, \odot} = 3.29$\,mag \citep{Blanton:2007}.

Firstly, all the vertical axes in \reffig{fig:tf} require a distance
estimate for each galaxy. As described in \citetalias{Williams:2009}, we use surface
brightness fluctuation (SBF) estimates where available (NGC~1381 from
\citealt{Jensen:2003} and NGC~1596 from \citealt{Tonry:2001}) and the
Virgo cluster distance where appropriate (NGC~4469 and NGC~4710,
\citealt{Mei:2007}). For all other galaxies we adopt the redshift
distance presented in the NASA/IPAC Extragalactic Database, corrected
for a Virgocentric flow model \citep{Mould:2000}. We assume a
\emph{Wilkinson Microwave Anisotropy Probe} five-year (WMAP5) cosmology
($H_0 = 70$\,km\,s$^{-1}$\,Mpc$^{-1}$, \citealt{Komatsu:2009}). Of
course, we avoid distance estimates derived from TFRs.

The total apparent magnitude at $K_S$-band is measured using the
analytically derived total light in the multi-Gaussian expansions
\citep{Emsellem:1994,Cappellari:2002} of our near-infrared images, as
described and tabulated in \citetalias{Williams:2009}. The total apparent
magnitude at $B$-band is taken from HYPERLEDA \citep{Paturel:2003}. We
apply corrections for Galactic ($A_\mathrm{G}$) and internal
($A_\mathrm{i}$) extinction in both bands using the Galactic dust maps
of \cite{Schlegel:1998} and the internal extinction correction presented
in \cite{Bottinelli:1995}, as tabulated in HYPERLEDA
\citep{Paturel:2003}, both of which we transform to the appropriate
bands using the extinction law of \cite{Cardelli:1989}, implying that
$A_\mathrm{G}(K_S)/A_G(B) = A_\mathrm{i}(K_S)/A_\mathrm{i}(B) = 0.084$.
The internal extinction correction is a function of both bulge-to-disc
ratio and morphological type. This correction is $\lesssim 0.1$\,mag at
$K_S$-band for our spirals and smaller still for the S0s.

With the exception of the Virgo galaxies and the two galaxies with SBF
distances, we adopt characteristic distance errors corresponding to a
200\,\kms{} uncertainty in the flow model \citep[e.g.][]{Tonry:2000},
and 0.05\,mag uncertainties in the apparent magnitudes. These
uncertainties completely overwhelm any error in the Galactic or internal
extinction at $K_S$-band. At $B$-band, we adopt uncertainties of
0.16\,mag in Galactic extinction \citep{Schlegel:1998} and 0.1\,mag in
the internal extinction. We emphasize, however, that internal
extinction corrections in edge-on spiral galaxies are both large and
uncertain. The $B$-band TFR is problematic with even favourably inclined
spirals, but with our edge-on sample the internal extinction correction
and its uncertainties may introduce significant systematic errors. The
$B$-band data are therefore included in this work merely to aid
comparison with previous work, and we do not make use of these results
in our interpretation.

The stellar mass, $M_\star$, is derived from the total absolute
$K_S$-band luminosity and the constant stellar mass-to-light
ratio, $(M/L)_{K_S}$, of each galaxy determined in
\citetalias{Williams:2009}. As
described in \refsec{sec:defined}, this uses dynamical methods that make
assumptions about the dark halo and stellar dynamics, but are not
vulnerable to zero-point or colour-dependent uncertainties due to the
initial mass function or stellar evolution. 

The dynamical mass, $M_\mathrm{dyn}$, is estimated using the total
absolute $K_S$-band luminosity and the \emph{dynamical} mass-to-light
ratio at $K_S$-band for each galaxy, also determined in
\citetalias{Williams:2009} in which it is referred to as
$(M/L)_{K_S,\mathrm{nohalo}}$. Unless there is no dark matter (or its
distribution closely follows that of the luminous matter), the dynamical
$(M/L)$ ratio is likely to be a quantity that varies significantly
within a galaxy, increasing toward the edge of the optical disc where
dark matter normally begins to dominate. The value we use is, however,
effectively an average for each galaxy within the radii constrained by
the stellar kinematic data, typically 2-3\,\reff (where \reff is the
projected $K_S$-band half-light radius), but dominated by the central
regions. Because of this, $M_\mathrm{dyn}$ is a somewhat ill-defined
quantity. 

An alternative approach would be to adopt the mass within the
three-dimensional half-light volume, $M_{1/2}$. We show how this can be
evaluated from observations or models of the circular velocity in disc
galaxies in Appendix~\ref{sec:mass}. We also show there that, in
practice, $2M_{1/2} = M_\mathrm{dyn}$ to within 13 per cent. Neither
choice is optimal but in the context of the present work, we can think
of no reason why one choice or the other would introduce systematic
biases that vary between spirals and S0s. The choice is therefore
somewhat arbitrary and has no affect on the discussion of the
relationship between spirals and S0s that follows where we use
$M_\mathrm{dyn}$. This follows precedent (it is the definition used by,
e.g., \citealt{Magorrian:1998}, \citealt{Haring:2004} and
\citealt{Cappellari:2006}) and avoids the possible concern that the
argument we present in \refsec{sec:massdisc} is circular, but we
emphasize that $M_{1/2}$ gives the same results and is a better defined
quantity.

For reference, we show in \reffig{fig:tf} the spiral TFRs determined
using global \hi{} line widths at $K$-band by TP00, at $K_S$-band by
\cite{Masters:2008} and at $B$-band by TP00 and \cite{Sakai:2000}. We do
this assuming that the global \hi{} line width, once corrected for
inclination, broadening and turbulent motion, is exactly twice the
rotational velocity. The zero points of the TP00 and \cite{Sakai:2000}
TFRs were determined using known Cepheid distances to 4 galaxies in the
case of TP00 at $K$-band (1 Sab, 2 Sb and 1 Sc; see the companion paper
by \citealt{Rothberg:2000}), 24 galaxies for TP00 at $B$-band (3 Sab, 14
Sb\,--\,Sc and 7 Scd\,--\,Sd) and 21 galaxies for \cite{Sakai:2000} (3
Sa\,--\,Sab, 16 Sb\,--\,Sc and 2 Scd). We position the
\cite{Masters:2008} $K_S$-band TFR on our plots using our adopted value
of $H_0$. The difference between the two $K$-band and two $B$-band
relations gives an idea of the uncertainty in the absolute location of
the late-type spiral TFR (the difference between the $K$-band filter
used by TP00 and the 2MASS $K_S$-band filter is not significant in this
context). These comparison spiral TFRs are unconstrained by observations
above $\log (v/\mathrm{km\ s^{-1}}) \approx 2.45$, while a third of our
sample is above this value.

Unlike \citetalias{Bedregal:2006}, we do not shift the zero-points of
the TP00 TFRs by $-0.207$\,mag. \citetalias{Bedregal:2006} did this to
make the value of $H_0$ implied by TP00's data
($77\pm8$\,\kms\,Mpc$^{-1}$) consistent with their adopted value
($70$\,\kms\,Mpc$^{-1}$). However, we believe that this correction was
in error, since the absolute locations of the TP00 TFRs were set by
independent Cepheid distances, which imply rather than assume a
particular value of $H_0$.

We also show an extrapolation to high mass systems of the baryonic TFR
presented in \cite{Stark:2009}, which was calibrated using gas-dominated
dwarf galaxies (hence minimizing uncertainties associated with the
stellar mass-to-light ratio). Neither of our mass estimates $M_\star$ or
$M_\mathrm{dyn}$ are designed to reproduce the baryonic mass
$M_\mathrm{bar}$ used by \cite{Stark:2009}. In practice, however, one
expects $M_\star \approx M_\mathrm{bar}$ to within a few per cent for
gas-poor S0s and to within perhaps ten per cent for giant Sa--Sb
spirals, while $M_\mathrm{dyn} \approx M_\mathrm{bar}$ to within a
factor of 2 for both S0s and spirals (dark matter is no more than 50 per
cent of the mass within the optical radius of disk galaxies, e.g.
\citealt{van-Albada:1986,Persic:1996,Palunas:2000,Cappellari:2006,Kassin:2006},
\citetalias{Williams:2009}). The comparison of our stellar and dynamical
mass TFRs to the baryonic TFR is therefore meaningful, and we note that
the \cite{Stark:2009} TFR is broadly consistent with our spirals. The
good agreement gives us confidence in our mass models.

It is crucial to emphasize that at no point in our analysis do we do
anything that might systematically affect the S0s in the sample
differently to the spirals. One slight but unavoidable observational
bias may be introduced by the presence of dust along the major axes. If
the slit is placed exactly on the major axis, then absorption by dust
may make it impossible to recover the full line-of-sight velocity
distributions (LOSVDs). Depending on the optical thickness, this may
truncate the low velocity wing or, in very dusty discs, the high
velocity peak of the LOSVDs. When Gaussians are fitted to these
distribution \citep{Chung:2004}, this could therefore increase or,
perhaps more likely, decrease the derived line-of-sight velocities. This
effect would propagate into the rotation curves from which \vmodel{} and
\vdrift{} are derived, and also into the mass-to-light ratios inferred
in \citetalias{Williams:2009}, from which $M_\star$ and $M_\mathrm{dyn}$
are derived. However, in cases where the dust absorption was strong, the
slit was simply shifted from the major axis by $\approx 2$ arcsec to
ensure the full LOSVD was sampled, and the objects with strong dust
lanes were selected to be very slightly away from perfectly edge-on
\citep{Bureau:1999}.

If the rotational velocity is a strong function of $z$, the height above
the equatorial plane in the cylindrical coordinate system of the galaxy
$(R, \phi, z)$, this intentional shift could itself bias the velocities
and inferred masses in dustier galaxies (i.e. spirals). In a future work
(Williams et al., in preparation), we will present pseudo-integral-field
data (i.e. a sparse velocity fields constructed from multiple long-slit
positions) for six of the galaxies in the present sample. In these data
we see no evidence that $\ud\,v/\ud\,|z| \ne 0$ in the disks of five of
the galaxies. At most, there is a gradient of
$-30$\,km\,s$^{-1}$\,kpc$^{-1}$ in the disk of NGC~7123. This is broadly
consistent with observations of planetary nebulae in the edge-on spiral
NGC~891 \citep{Merrifield:2010} and of extraplanar gas (e.g. NGC~891,
\citealt{Heald:2007}, \citealt{Oosterloo:2007}; NGC~4302 and NGC~5775,
\citealt{Heald:2007}; the Milky Way, \citealt{Levine:2008}), which find
a typical falloff in rotation $\ud v/\ud |z| \approx
-20$\,km\,s$^{-1}$\,kpc$^{-1}$ In the worst case scenario, in which all
our spirals (and none of the S0s) suffered from absorption strong enough
to require shifting the slit, the spiral velocities would typically be
biased by $\lesssim -0.02$\,dex. We estimate that the overall effect on
our sample of spirals, most of which did not require this slit shift,
may be to bias their velocities by up to $\approx -0.01$\,dex. In
practice, the slit was 1.8 arcsec wide and rather difficult to align
precisely for both spirals and S0s. This imprecision, which introduces
velocity scatter in both samples, likely overwhelms any systematic
velocity falloff in the spirals due to intentionally shifting the slit.

\section{Fitting procedure}
\label{sec:fitting}

For each measure of luminosity or mass as a function of $x =
\log(\vmodel/\mathrm{km\ s^{-1}})$, we simultaneously fit a straight
line of the form $y = a(x - 2.4) + b$ to the spiral galaxies and $y =
a(x - 2.4) + b + \delta$ to the S0s. The two straight lines are
therefore constrained to have the same slope, $a$, but they have
zero-points which differ by $\delta$. In separate fits to the two
samples, we find that their slopes are consistent within the
uncertainties. Constraining them to be equal therefore significantly
simplifies the interpretation of our results in terms of zero-point
evolution. The $x = 2.4$ reference value is defined to minimize the
covariance between errors in $a$ and $b$ \citep[e.g.][]{Tremaine:2002}.
Its choice does not affect the results of this work, in which were are
mainly interested in the zero point offset $\delta$.

To find the optimal combination of $a$, $b$ and $\delta$ for a given
scatter, we define and minimize the figure of merit
\begin{equation}
\label{eqn:chi2}
\begin{split}
\chi^2 &\equiv \sum^{N_\mathrm{Sa\,Sb}}_{i = 0} \frac{1}{\sigma_i^2} \{y_i - [a(x_i - 2.4) + b]\}^2\\
&\quad + \sum^{N_\mathrm{S0}}_{i = 0} \frac{1}{\sigma_i^2} \{y_i -
[a(x_i - 2.4) + b + \delta]\}^2,
\end{split}
\end{equation}
where $N_\mathrm{Sa\,Sb}$ is the number of spirals, $N_\mathrm{S0}$ the number
of S0s and 
\begin{equation}
\sigma_i^2 \equiv \sigma_{y,i}^2 + a^2 \sigma_{x,i}^2 + \sigma_\mathrm{int}^2.
\end{equation}
The extra scatter $\sigma_\mathrm{int}$ is iteratively adjusted
to ensure that
\begin{equation}
\chi_\mathrm{red}^2 \equiv \frac{\chi^2}{(N_{\rm Sa,Sb} + N_{\rm S0} -
3)} \approx 1.
\end{equation}
If the observational errors for each galaxy $\sigma_{x,i}$ and
$\sigma_{y,i}$ are well-estimated then $\sigma_\mathrm{int}$ is the
intrinsic, astrophysical scatter in the TFR. 

We also define a total rms scatter $\sigma_\mathrm{tot}$, essentially a
weighted mean of the scatter of the data about the fits (and therefore
in the same units as the $y$-axis): 
\begin{align}
\label{eqn:totalscatter}
\begin{split}
\sigma_{\rm tot}^2 &\equiv 
\Bigg(\sum^{N_\mathrm{Sa\,Sb}}_{i = 0} \frac{1}{\sigma_i^2}\{y_i - [a(x_i - 2.4) + b]\}^2 \\
&\quad + \sum^{N_\mathrm{S0}}_{i = 0} \frac{1}{\sigma_i^2}\{y_i - [a(x_i - 2.4) + b + \delta]\}^2\Bigg)
\Bigg/\sum^{N_\mathrm{gal}}_{i = 0} \frac{1}{\sigma_i^2}
\end{split}\\
&=\frac{\chi^2}{\sum_{i=0}^{N_\mathrm{gal}} 1/\sigma_i^2},
\end{align}
where $N_\mathrm{gal}$ is the total number of both spiral and S0
galaxies. Comparing $\sigma_\mathrm{tot}$ with $\sigma_{\rm int}$ gives
an idea of what fraction of the scatter is intrinsic and what fraction
is due to measurement errors. We minimize equation (\ref{eqn:chi2}) using
the \textsc{mpfit} package \citep{Markwardt:2008}.

Following previous analyses of the Tully-Fisher relation, we also fit
the `inverse' relation, i.e. we regress the observed rotational
velocities rather than the luminosities or masses onto the model (see,
e.g., TP00, \citealt{Verheijen:2001}, \citealt{Pizagno:2007} and
references therein).
We rewrite the spiral and S0 straight lines as $x = Ay + B + 2.4$ and $x
= Ay + B + 2.4 + \Delta$ and minimize
\begin{equation}
\label{eqn:chi2inv}
\begin{split}
\chi^2_\mathrm{inv} &\equiv \sum^{N_\mathrm{Sa\,Sb}}_{i = 0} \frac{1}{\varsigma_i^{2}}[x_i - (Ay_i + B + 2.4)]^2 \\ 
&\quad + \sum^{N_\mathrm{S0}}_{i = 0} \frac{1}{\varsigma_i^{2}}[x_i - (Ay_i + B + \Delta + 2.4)]^2,
\end{split}
\end{equation}
where 
\begin{equation}
\varsigma_i^{2} = \sigma_{x,i}^2 + A^{2} \sigma_{y,i}^2 + \varsigma_\mathrm{int}^{2} 
\end{equation}
and
\begin{equation}
\varsigma^2_\mathrm{tot} = \frac{\chi_\mathrm{inv}^2}{\sum_{i=0}^{N_\mathrm{gal}} 1/\varsigma_i^2}.
\end{equation}
To compare these values to those determined using the forward relation,
we use the fact that a best-fitting inverse relation has an equivalent
forward slope $1/A$, zero-point $-B/A$, offset $-\Delta/A$ and $y$-axis
intrinsic and total scatters $A\varsigma_\mathrm{int}$ and
$A\varsigma_\mathrm{tot}$. Our implementation of the fitting procedure
described above is publicly
available\footnote{\url{http://purl.org/mike/mpfitexy}}.

Forward and inverse fitting is only symmetric if there is no intrinsic
scatter ($\sigma_\mathrm{int} = \varsigma_\mathrm{int} = 0)$
\citep{Tremaine:2002}. This is not generally the case and the slopes can
be very different. In analyses of the TFR with observed data, it has
been shown that there is a significant bias in the slope of the forward
line of best fit \citep{Willick:1994}, which is why the inverse
relationship is usually preferred. The figures in this paper use this
inverse fitting approach, but to aid comparison with future work, we
present both forward and inverse fits in Table~\ref{tab:params}. The
slopes of the forward and inverse fits to our data are indeed
discrepant, often to the extent that they do not lie within each other's
error bars (especially at $B$-band). We are fortunate, however, that the
choice of whether to use the forward or inverse relationship does not
affect our conclusions, which depend entirely on the zero-point offset
of the S0 TFR and the total scatter. 

\section{Discussion}
\label{sec:discussion}

\subsection{The TFR zero-point offset at $K_S$-band and $B$-band}
\label{sec:lumdisc}

\begin{table*}
\begin{minipage}{12cm}
\caption{Parameters of the best fitting TFRs, where $x =
\log(\vmodel/\mathrm{km\,s}^{-1})$, $y = a(x - 2.4) + b$
for spirals and $y = a(x - 2.4) + b + \delta$ for S0s.}
\begin{tabular*}{\textwidth}{@{\extracolsep{\fill}}lr@{\extracolsep{0cm}}@{$\pm$}l@{\extracolsep{\fill}}r@{\extracolsep{0cm}}@{$\pm$}l@{\extracolsep{\fill}}r@{\extracolsep{0cm}}@{$\pm$}l@{\extracolsep{\fill}}cc}
\noalign{\smallskip}
\hline
$y$ & \multicolumn{2}{c}{$a$} & \multicolumn{2}{c}{$b$}  &  \multicolumn{2}{c}{$\delta$} & $\sigma_{\rm int}$ & $\sigma_{\rm tot}$ \\
\hline
\noalign{\smallskip}
\multicolumn{9}{c}{Forward regression} \\
\noalign{\smallskip}
                     $M_{K_S}$ & $ -8.15$&$ 0.76$ & $  -24.53$&$ 0.11$ & $ 0.55$&$ 0.15$ & 0.27 & 0.37 \\ 
                         $M_B$ & $ -7.67$&$ 1.02$ & $  -20.82$&$ 0.14$ & $ 0.73$&$ 0.20$ & 0.39 & 0.50 \\ 
       $\log(M_\star/M_\odot)$ & $  3.45$&$ 0.26$ & $   11.14$&$ 0.04$ & $-0.22$&$ 0.05$ & 0.06 & 0.13 \\ 
$\log(M_\mathrm{dyn}/M_\odot)$ & $  3.34$&$ 0.27$ & $   11.23$&$ 0.04$ & $-0.20$&$ 0.05$ & 0.07 & 0.13 \\ 
\noalign{\smallskip}
\multicolumn{9}{c}{Inverse regression} \\
\noalign{\smallskip}
                     $M_{K_S}$ & $ -9.13$&$ 0.87$ & $  -24.49$&$ 0.12$ & $ 0.51$&$ 0.16$ & 0.28 & 0.39 \\ 
                         $M_B$ & $ -9.81$&$ 1.26$ & $  -20.76$&$ 0.16$ & $ 0.64$&$ 0.23$ & 0.44 & 0.56 \\ 
       $\log(M_\star/M_\odot)$ & $  3.57$&$ 0.27$ & $   11.13$&$ 0.04$ & $-0.21$&$ 0.05$ & 0.06 & 0.13 \\ 
$\log(M_\mathrm{dyn}/M_\odot)$ & $  3.52$&$ 0.28$ & $   11.22$&$ 0.04$ & $-0.19$&$ 0.05$ & 0.08 & 0.13 \\ 
\noalign{\smallskip}
\hline
\noalign{\smallskip}
\end{tabular*}
\begin{minipage}{\textwidth}
\emph{Notes}. 
$\sigma_\mathrm{int}$ is the intrinsic scatter required to yield
$\chi_\mathrm{red} = 1.00\pm0.01$ and $\sigma_\mathrm{tot}$ is the total
scatter (see equation~\ref{eqn:totalscatter}). $a$
is in units of $y/x$, all other quantities are in units of $y$, i.e. magnitudes
or dex of solar masses. \end{minipage}
\label{tab:params}
\end{minipage}
\end{table*}

While the TFR zero point and offset are formally sensitive to the choice
between forward or inverse regression, in practice they are almost
unchanged by this choice. In the discussion that follows, we adopt the
means of the forward and inverse values and their errors.

We detect a statistically significant difference between the zero-points
of the TFRs of spiral and S0 galaxies (see Table~\ref{tab:params}). At a
given \vmodel, local S0s are $0.53\pm0.15$ mag fainter at $K_S$-band and
$0.68\pm0.22$ mag fainter at $B$-band than local spirals. The systematic
uncertainty is due to the slit positioning issue described at the end of
\refsec{sec:data}. We emphasize again that the $B$-band offset we
measure is dubious because of the very uncertain (and very large)
internal extinction corrections at optical wavelengths in edge-on
spirals. The $K_S$-band value is, however, robust. 

We estimate that the intentional misalignment of the slit in the
dustiest galaxies may introduce a bias of up to $\approx -0.01$\,dex in
the observed velocities of spirals (see the end of \refsec{sec:data}).
In principle, this may bias our measurement of the $K_S$-band offset
to be too large by $\approx 0.07$\,mag. This is less than uncertainty in
the intercept of the spiral TFR ($\pm0.10$\,mag) or the offset of S0 TFR
($\pm0.15$\,mag), so we do not discuss it further.

We restate that our fiducial TFRs use \vmodel{} as the measure of
rotation. We emphasize, however, that the \emph{offsets} we measure
between the spiral and S0 TFRs are completely insensitive to this
choice. Using \vdrift{}, there is an offset at $K_S$-band of
$0.45\pm0.18$\,mag. Using \vgas{}, there is an offset of
$0.48\pm0.24$\,mag. Unless all three of these measures of rotation are
flawed then our principal result, the measurement of the offset of the
S0 TFR, is reliable. (We neglect \vhi{} because it is based on spatially
unresolved data and sometimes unphysically discrepant from the other
measures of rotation. See \refsec{sec:comparison}.) 

Since high redshift spirals are thought to be the progenitors of local
S0s, we would really like to compare our local S0 TFR to the high
redshift spiral TFR. As we emphasize in \refsec{sec:rotation}, however,
the comparison of TFRs derived using different measures of rotation is
uncertain at best, so the zero-points of high-redshift TFRs cannot be
compared to our local results in a meaningful way. In addition to the
uncertainties in comparing the velocities of different tracers in the
local universe, the move to high-$z$ raises the possibility that the
intrinsic \emph{shapes} of rotation curves have changed, which we cannot
rule out. Provided, however, that these differences have been correctly
accounted for by other authors, the relative shifts between the local
and high-$z$ TFRs should be fairly insensitive to the measure of
rotation used. We therefore make that unavoidable assumption in order to
proceed.

\cite{Conselice:2005}, \cite{Flores:2006} and \cite{Kassin:2007} find no
evidence for evolution in either the slope or zero-point of the $K$-band
spiral TFR from the local universe to $z \approx 1$. However,
\cite{Puech:2008} detect a dimming of 0.66\,mag at $K_S$-band from $z =
0$ to $z \approx 0.6$. We first discuss the possibility that the spiral
TFR has not evolved with redshift. If this is correct, our results imply
that local S0s are $0.53\pm0.15$\,mag fainter at $K_S$-band for a given
rotational velocity than their presumed spiral progenitors.

To get a feeling for how long such a fading would take, we follow
\citetalias{Bedregal:2006} by
using the \cite{Bruzual:2003} stellar population synthesis code to
assign an approximate timescale to such a fading under various star
formation scenarios. We assume solar metallicity and a Chabrier initial
mass function \citep{Chabrier:2003} and use an updated prescription for
the thermally-pulsing asymptotic giant branch (\citealt{Marigo:2007};
Charlot 2009, private communication). Following a constant star formation
episode lasting 5\,Gyr, the synthetic stellar population takes
$0.9^{+0.4}_{-0.5}$\,Gyr to fade by $0.53\pm0.15$\,mag at $K_S$-band and
$0.2^{+0.2}_{-0.1}$\,Gyr to fade by $0.67\pm0.21$\,mag at $B$-band. If
we add an instantaneous `last gasp' burst of star formation comprising
10\% of the mass of the galaxy at the end of the 5\,Gyr episode
\citep{Bedregal:2008}, the timescales increase to
$1.4^{+0.4}_{-0.2}$\,Gyr at $K_S$-band and $0.5^{+0.2}_{-0.1}$\,Gyr at
$B$-band.

For plausible star formation histories, the timescales implied by the
offsets at the two wavelengths are inconsistent. This is perhaps not
surprising given the uncertainty in the dust corrections at $B$-band.
However, the $K_S$-band offset is far less susceptible to this possible
systematic error, and this measurement implies an uncomfortably short
timescale since the truncation of star formation of $\lesssim 1.4$\,Gyr,
corresponding to a redshift  $z \approx 0.1 $. As noted by
\citetalias{Bedregal:2006}, it would
be a surprising coincidence if we were living in an era so soon after
the S0s in our sample transformed from spirals. Observationally, this
possibility appears to be ruled out by high-redshift observations: while
\cite{Dressler:1997} show that the S0 fraction has risen at the expense
of spirals since $z \approx 0.5$ (5\,Gyr ago), \cite{Fasano:2000} find
that the present relative fraction of S0s and spirals was largely
already in place in clusters at $z \approx 0.2$ (2.5\,Gyr ago) and there
was no shortage of S0s in groups at $z \approx 0.4$ (4\,Gyr ago,
\citealt{Wilman:2009}). In the absence of strong environmental
processes, we naively expect the transformation of field spirals to S0s
to be gradual, and therefore that present day field S0s would
need to have begun their transformation even earlier. We therefore
conclude that a simple scenario of passive fading of spirals into S0s is
inconsistent with the TFRs of our sample and the evolution of the
S0/spiral fraction with redshift. This interpretation of the offset of
the S0 TFR is consistent with that of \cite{Neistein:1999}. With limited
data they  measured a very similar S0 TFR offset to us, but with much
larger uncertainties, and with respect to a line width-based TFR (a
comparison we argue is flawed in \refsec{sec:rotation}).

We now discuss the possibility that the zero-point of the TFR was
fainter at earlier times. If the $K$-band spiral TFR has evolved with
redshift and was, for example, 0.66\,mag fainter at $z \approx 0.6$ than
the local relation (as found by \citealt{Puech:2008}), then present-day
S0s are approximately as luminous at a given rotational velocity as
spirals were 6\,Gyr ago, i.e. there is no evidence for any evolution in
the velocity--luminosity plane between local S0s and their presumed high
redshift progenitors. A simple star formation truncation scenario is
then even harder to reconcile with our results.

The above analysis is based on the assumption that an S0 of a given
rotational velocity used to have the luminosity of a high-redshift
spiral of the same rotational velocity. From this assumption we have
gone on to consider in isolation the evolution of a single broadband
colour. The parameter space of possible star formation histories is,
however, highly degenerate and this is far from the ideal way to
constrain star formation histories. It is nevertheless clear from our
sample that something other than simple fading beginning at the
redshifts at which S0s are first observed (and without subsequent star
formation), is transforming spirals to S0s. The evidence for recent (and
even ongoing) star formation in local S0s from UV
\citep{Kaviraj:2007,Jeong:2009} and mid-infrared
\citep{Temi:2009,Temi:2009a,Shapiro:2010} observations is in fact
strong.

\subsection{Stellar and dynamical mass}
\label{sec:massdisc}

We also detect a significant difference between the zero points of the
S0 and spiral stellar mass TFRs. The S0s have around 0.2\,dex less
stellar mass for a given rotational velocity than the spirals. Because
of the relative gas richness of spirals, the offset in the baryonic
relation \citep[i.e. stellar mass $+$ gas mass, e.g.][]{McGaugh:2000} is
likely larger and certainly no smaller. Assuming that \vmodel{} traces
the enclosed dynamical mass equally well for S0s and spirals, this might
naively suggest that S0s have less stellar mass per dynamical mass
within 2--3\,\reff{} (the extent of our kinematic data), and therefore
contain fractionally more dark matter than spirals. This is generally
consistent with the idea that S0s are older systems that are found in
denser environments at the bottom of deeper potential wells. If this is
indeed the case, using dynamical rather than stellar mass should
eliminate the offset. The dynamical mass TFR is essentially a plot of a
one measure of dynamical mass (from a mass model) against another (the
observed rotational velocity), and since gravity applies equally to
spirals and S0s there should be no difference.

Having said that, the difference between the spiral and S0 zero points
persists almost unchanged in the dynamical mass TFR (see
Table~\ref{tab:params}). Our dynamical modelling approach does
necessarily make approximations and assumptions, but it is unlikely that
the models could be systematically biased by morphological type. Our
stellar and dynamical mass spiral TFRs are also consistent with the
reliably calibrated baryonic TFR of \cite{Stark:2009}. This raises the
possibility that S0s are systematically smaller or more concentrated
than spirals. Recall that $M_\mathrm{dyn} \propto v^2 R$, where $R$ is
some characteristic radius. If $R$ is systematically smaller in S0s at a
given rotational velocity, or if the dynamical mass is more concentrated
such that more mass is enclosed within a given characteristic radius,
then that could explain the offset in the dynamical mass TFR. In other
words, spirals and S0s do not form a homologous family. In this
scenario, the offset of the S0 dynamical mass TFR should not be thought
of as a -0.18\,dex offset in mass (as presented in
Table~\ref{tab:params}), but, because of the slope of the TFR, a
+0.05\,dex offset in velocity. This could be due either to S0s being
approximately 80\% of the size of spirals of the same rotational
velocity, or their dynamical mass being distributed more compactly such
that the characteristic radius probed by measures like \vmodel{} and
\vdrift{} contains approximately 25\% more dynamical mass. Again, since
S0s are thought to be older, more evolved systems with bulge-to-disc
ratios that are larger on average than those of spirals, this is perhaps
consistent with naive expectations, but is it true for our sample? In
the spirit of \cite{Courteau:1999} and \cite {Courteau:2007}, the
possibility that the luminous component of the dynamical mass is smaller
or more concentrated in S0s at a given velocity can be tested directly
by measuring the size of our galaxies. 

We define $R_\mathrm{d}$ as described in \refsec{sec:defined} (where it
is used as a parameter of the asymmetric drift correction) and \reff{}
as the semi-major axis of the ellipse containing half the light at
$K_S$-band (see section 4.2 of \citetalias{Williams:2009} for a discussion of
alternative definitions). In an edge-on galaxy with a boxy bulge, such
as those in our sample, the radial surface brightness profile often
contains a plateau or even a secondary maximum \citep{Bureau:2006}. This
means the disk scale length is not well defined. Measurement of \reff{}
depends sensitively on the ellipticity of the aperture used
\citepalias{Williams:2009}. Moreover, our data probe a relatively small range in
\vmodel{}. Together, these difficulties mean that the constraints we can
place on the parameters of the size--velocity relation are very weak. We
recover a correlation $R_\mathrm{d} \propto \reff \propto
\vmodel^\gamma$, where $\gamma \approx 1.5\pm0.5$. Significantly, we
also detect some evidence of a systematic difference between the zero
points of the size--velocity relations of S0s and spirals. Using the
simultaneous line fitting approach described in \refsec{sec:fitting}, we
find that S0s are on average $0.15\pm0.10$\,dex smaller than spirals at
a given \vmodel{}. This result is of weak statistical significance, but
is consistent with our interpretation. We see no significant difference
between the mean compactness of the light distribution, $C_{82}$, of the
S0s and spirals. $C_{82} \equiv R_{80}/\reff$, where $R_{80}$ is the
semi-major axis of the ellipse containing 80\% of the light.

The large sample of \cite{Courteau:2007} does not include S0s, but
extends from Sa to Sc and is made up of galaxies that are more
favourably inclined for measurements of size. Recalling that $K_S$-band
luminosity is a good proxy for stellar mass \citep[e.g.][]{Bell:2001},
which is a reasonable proxy for dynamical mass in the optical parts of
disc galaxies
\citep[e.g.][]{van-Albada:1986,Englmaier:1999,Palunas:2000,Weiner:2001},
\cite{Courteau:2007} find a morphological trend in the sense required to
explain our result: earlier-type spirals are smaller for a given
$K_S$-band luminosity than later-types (see their table 3 and figure
11). At a characteristic $K_S$-band luminosity of $10^{11} L_\odot$,
they find that Sas are around 80\% the size of Sbs, which are in turn
around 80\% the size of Scs. If this trend extends to S0s, then its
sense and magnitude are consistent with the offset we see in our
dynamical mass TFR. 

There is also the possibility that the dark halos (rather than the
baryonic mass distribution) of S0s are smaller or more concentrated than
those of spirals. Unfortunately we are unable to quantify this with our
kinematic data, which are not radially extended enough to break the well
known degeneracy between halo mass and concentration
\citep{van-Albada:1985}. However, the inability of the dynamical models
presented in \citetalias{Williams:2009} to quantify the amount of halo contraction has no
effect on the results presented here. This is because, even if one
assumes different dark halo shapes, the parameters of the best fitting
models always conspire to produce an almost unchanging \vc{} profile in
the region constrained by the data. This is ultimately the reason why
the degeneracy exists in the first place. A quantitative illustration of
the robustness of the circular velocity profiles recovered from
dynamical models to the degeneracy in the halo concentration is given
in, e.g., fig. 9 of \cite{Thomas:2009}. 

The interpretation of the systematic offset of the S0 TFR to higher
velocities due to a more concentrated mass distribution or smaller size
than spirals is consistent with the discussion of the shapes of very
extended gas rotation curves in \cite{Noordermeer:2007a} and
\cite{Noordermeer:2007}. Their kinematic data extend well beyond the
optical discs of the galaxies, and allow them to study the asymptotic
behaviour of the rotation curves. They find that the earlier galaxies in
their sample of S0--Sab galaxies have rotation curves that reach a
maximum that is greater than their asymptotic velocity and occurs at
smaller radii. They argue that this behaviour can be understood in terms
of the larger bulge-to-disc ratios of the earlier-type galaxies in their
sample. The scatter of their earlier-type galaxies away from the TFR of
the late type galaxies is reduced by using the asymptotic velocity,
which again suggests the Tully-Fisher relation is a manifestation of a
close connection between galaxy luminosity (or mass) and halo mass.

If the size/concentration argument discussed above is indeed the
explanation for the difference between the zero points of the S0 and
spiral TFRs as a function of dynamical mass, then the offsets we observe
as a function of luminosity (and discuss extensively in
\refsec{sec:lumdisc} in the context of fading relative to the
high-redshift spiral TFR) are in fact due to differences in the
distributions of dynamical mass in spirals and S0s. The offset between
spirals and S0s in the Tully-Fisher (luminosity--velocity) relation may
therefore not be a property of their stellar populations as argued in
\citetalias{Bedregal:2006}, but rather the result of more fundamental offset between spirals
and S0s in the mass--size (or luminosity--size) projections of the
luminosity--size--velocity plane occupied by disc galaxies. Models of S0
formation in which S0s are end points of spiral evolution should
consider this possibility.

\subsection{Scatter of the TFRs}

The $B$-band luminosities  we use, which are drawn from HYPERLEDA, are
less accurate than our $K_S$-band magnitudes, which are based on a
relatively recent, homogeneous set of observations. The extinction
corrections at $B$-band are also particularly large and uncertain for
edge-on galaxies like ours. These uncertainties are probably not
reflected properly in our error estimates, and this may contribute to
the fact that the scatter of the TFR at $B$-band ($0.53$\,mag,
of which $0.42$\,mag is intrinsic rather than observational) is larger
than that at $K_S$-band ($0.38$\,mag, of which $0.28$\,mag is
intrinsic). 

These scatters are broadly consistent with dedicated studies of the
spiral TFR scatter \citep[e.g.][]{Pizagno:2007}, which is perhaps
somewhat surprising given the expected poor quality of our distance
estimates and the importance and uncertainty of internal extinction in
edge-on spirals. It should be borne in mind, however, that such studies
fit a single line to a range of morphological types, while we separately
fit two lines to two sub-samples, which reduces the measured total and
intrinsic scatters. If a single line is fitted to the composite sample
of spirals and S0s then the scatter increases to 0.50\,mag, of which
0.41\,mag is intrinsic.

Because our sample is drawn from a wide range of environments, the S0s
may have been transformed from spirals via multiple channels (or one
channel which began at different times in different galaxies). If so,
then the extent to which S0s have left the spiral TFR should vary, and
their scatter about the TFR should be larger than that of spirals. Our
fitting method implicitly assumes that spirals and S0s have the same
intrinsic scatter. However, our analysis also shows that there is no
evidence of any difference between the intrinsic or total scatter of
spiral galaxies about their line of best fit and that of S0s about
theirs. The absence of increased scatter in the S0 TFR is problematic
for all explanations of the offset (passive fading, environmental
stripping, minor mergers and the possible non-homology described in
\refsec{sec:massdisc}).

There is no significant difference between the total scatters of the
stellar mass TFRs (0.13\,dex, i.e. 35 per cent) and the scatter at
$K_S$-band (0.32\,mag, i.e 34 per cent in luminosity). This is probably
due to the relative constancy of mass-to-light ratios in the
near-infrared, from which the masses are derived. The difficulty of
assigning errors to the total masses of the dynamical models makes
drawing strong conclusions from the intrinsic scatter of the stellar
mass or dynamical mass TFRs difficult.

\subsection{Classification of edge-on disc galaxies}
\label{sec:classification}

The ability to reproducibly and correctly distinguish between spiral and
S0 galaxies is crucial for this work. Although spirals arms are not
visible in edge-on disc galaxies, the classification of edge-on galaxies
as spirals or S0s is clearly specified by the presence of extended dust
\citep[see, e.g.,][]{Hubble:1936,Sandage:1961,de-Vaucouleurs:1991}. It is
clear, therefore, that the classification of edge-on galaxies is
objective and reproducible. 

Although the classification of edge-on disc galaxies is reproducible,
would a galaxy be classified as the same type if viewed face-on? To
answer this concern, we note that if the effects of dust are visible in
a face-on spiral then, due to line-of-sight projection, they would be
even clearer if the same galaxy were viewed edge-on. For the simple
distinction between spiral and S0 galaxies (rather than more detailed
division into the Sa--Sd classes based on bulge size and the tightness of
spiral arms), edge-on orientation is thus arguably the
optimal viewing angle. This is also true of the distinction between S0s
and ellipticals, where a face-on orientation makes it extremely difficult
to detect the featureless stellar disc of an S0. Having said that, it is
clear that if an edge-on galaxy has just enough dust to be classified as
a spiral, it would probably (but perhaps wrongly) be classified as an S0
if viewed face-on. This will occur only for limiting cases, however, and
it is clear that the majority of edge-on classifications are correct.

\begin{figure}
\includegraphics[width=8.4cm]{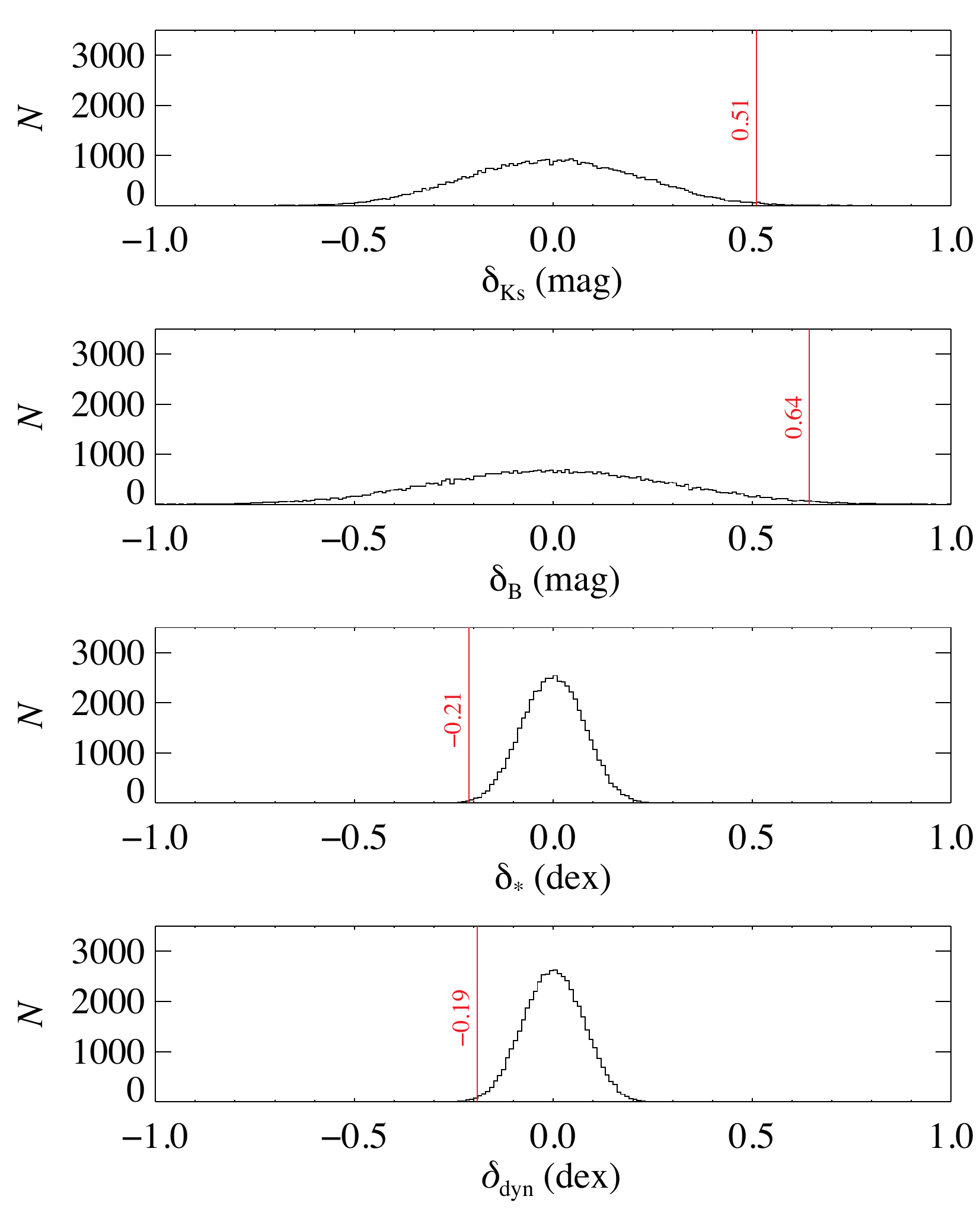}
\caption{Histograms showing the TFR zero-point offset $\delta$ between two 
sub-samples drawn randomly (with no regard to classification) from our
full sample of 28 galaxies. The inverse regression fitting procedure was
repeated 50,000 times to construct the histograms. From top to bottom,
we show the offsets for the $K_S$-band, $B$-band, stellar mass and
dynamical mass TFRs as a function of \vmodel. The vertical lines show
the value of the zero-point offset measured for the spiral and S0
sub-samples used here.}
\label{fig:randsample}
\end{figure}

Having established that a meaningful morphological classification of
edge-on disc galaxies is possible, we now ask whether this simple
spiral--S0 division based on the presence of dust is dynamically
significant. To test this, we randomly divided the complete sample of 28
galaxies into two sub-samples of 14 galaxies each and then measured the
offset $\delta$ between the zero points of the TFRs of the two random
sub-samples. We repeated this procedure 50,000 times to build up an idea
of how the offset between randomly selected sub-samples is distributed,
and how exceptional the adopted morphological classification is. This
idea is a simplification of the non-parametric test presented in the
context of comparing local and high-redshift TFRs by \cite{Koen:2009}.
Our results are presented in \reffig{fig:randsample}. It is clear that
the offsets observed between our S0s and spirals cannot to be due to
chance. In all cases ($y = M_{K_S}$, $M_B$, $M_\star$ or
$M_\mathrm{dyn}$, and for both forward and inverse regressions), the
observed offset is at least three standard deviations away from the mean
of the offsets between randomly selected sub-samples. This result
clearly demonstrates that a purely morphological classification of
edge-on early-type disc galaxies, based only on the presence of dust,
divides them into two truly kinematically distinct classes.

\subsection{Comparison to previous work}
\label{sec:previous}

The absolute location of our S0 TFR is almost identical to that measured
in \citetalias{Bedregal:2006} at both $K_S$ and $B$-band. They used \vdrift{} rather than
\vmodel, so this is a reflection of the consistency of these two
measures of rotation. If \citetalias{Bedregal:2006} could have used our spiral TFR as a
reference zero point, they would have measured the same offset as us.
However, they used the TFRs of TP00. As a result, our offsets are in the
same sense (S0s are fainter than spirals) but significantly smaller in
size than those measured in \citetalias{Bedregal:2006} (1.0$\pm$0.4\,mag at $K_S$,
1.6$\pm$0.4\,mag at $B$-band, where we have removed their correction of
the TP00 zero point). Because the zero-point offsets we measure are
smaller than those in \citetalias{Bedregal:2006}, the timescale for a simple synthetic
stellar population with a plausible star formation history to fade is
shorter too. There are at least two reasons for the larger offsets
measured in \citetalias{Bedregal:2006}, which could combine to account for the entirety of
the difference, making our results consistent.

The first is the possible intrinsic difference between the zero point of
the spiral TFR we compare our S0s to in the present work, and the spiral
TFR used in \citetalias{Bedregal:2006}. Our spiral TFR is constructed using Sa and Sb
galaxies, while that of TP00, which is used by
\citetalias{Bedregal:2006}, is calibrated
using later types (mostly Sbc and later). As shown by
\cite{Masters:2008}, in the velocity domain of the present work
($\left<v\right> \approx 250$\,\kms), Sa spirals are $\approx 0.4$\,mag
fainter than Sb and Sc spirals of the same rotational velocity. On the
other hand, \cite{Courteau:2007} estimate that Sas are just 0.1\,mag
fainter than Sbs. Indeed, with our much smaller statistics, we see no
evidence of a zero point variation between the Sa and Sb galaxies in our
sample.

The second possible effect is the systematic bias introduced by
comparing an S0 TFR derived from stellar kinematics to a spiral TFR
derived from global \hi{} line widths. Comparing the methods shows that
Sa--Sb galaxies are measured to be rotating $\approx 0.08$\,dex slower
in \hi{} than in their stellar kinematics (see \reffig{fig:vall}). At a
typical $K_S$-band TFR slope of $\approx 8$, this corresponds to a
magnitude shift of $\approx 0.6$\,mag. However, as we emphasized in
\refsec{sec:rotation}, applying a constant shift to correct for the
difference between stellar and \hi{} rotation measures is dubious because
of the large scatter in this difference (admittedly less so for
spirals).

It is easy to see how morphological differences between galaxies in the
reference spiral samples could combine with the systematic bias
introduced by the use of different measures of rotation to account for
the discrepancy between the fiducial offset presented here
($0.53\pm0.15$\,mag) and that in \citetalias{Bedregal:2006} ($1.0\pm0.4$\,mag). 

Another possible source of bias in the offset measured in
\citetalias{Bedregal:2006} is the
fact that the TP00 spiral TFR is constrained by observations of a sample
of relatively low mass spirals with $v \approx 150$\,km\,s$^{-1}$, while
\citetalias{Bedregal:2006}'s S0s have a mean $v \approx 250$\,km\,s$^{-1}$. If the slope of
the TFR changes with mass, as has been suggested
\citep[e.g.][]{Peletier:1993,Noordermeer:2007}, then the fact that the
S0s and spirals in our sample lie in the same velocity domain is a
crucial advantage.

In any case, because the rotation measure bias is unphysical and
unrelated to the formation and evolution of S0s, we argue that the
method we have employed here should be preferred if the goal is to
determine how much fainter local S0s are than local spirals.

Finally, we note that \citetalias{Bedregal:2006} found a much larger
scatter in their S0 TFR than we do here, $\approx 0.9$\,mag at
$K_S$-band. This may be explained by the relatively homogeneous nature
of our data compared to the multiple sources from which
\citetalias{Bedregal:2006} drew theirs, and perhaps to the larger
velocity and luminosity ranges their data probe.

\section{Conclusions}
\label{sec:conclusion}

We have demonstrated that the comparison of Tully-Fisher relations
derived from \hi{} line widths, ionized gas PVDs, stellar
kinematics corrected for asymmetric drift or the circular velocity
profiles of dynamical models is influenced by systematic and uncertain
biases introduced by the different measures of rotation used. We
therefore argue that to constrain the relative locations of the spiral
and S0 TFRs, the same tracer and methods must be used for both samples.
In practice, because of the paucity of extended, undisturbed \hi{} and
ionised gas in S0s, this means one must use stellar kinematics or
dynamical models.

In this work we used the circular velocity profiles of mass models to
construct TFRs for 14 Sa--Sb spirals and 14 S0s, eliminating the biases
introduced by mixing measures of rotation. The circular velocity curves
are those of models derived by solving the Jeans equations for mass
models comprising an axisymmetric stellar component and a spherical NFW
halo \citepalias{Williams:2009}. The parameters of the models are
constrained by observed long-slit major-axis stellar kinematics
\citep{Chung:2004}. We characterized the circular velocity profile for
each galaxy by single numbers, \vmodel{}, by taking the average in the
flat part of the observed rotation curve. 

By simultaneously fitting two offset relations with a common slope to
this spirals and S0s, we find that S0s are systematically
$0.53\pm0.15$\,mag fainter at $K_S$-band than local Sa--Sb spirals of
the same rotational velocity. This measurement is almost identical if we
use estimates of the rotational velocity derived from ionized gas PVDs
or stellar kinematics corrected for asymmetric drift.

If the high-redshift spiral TFR has the same zero point as the local
spiral TFR, this is inconsistent with the observed evolution of the
spiral/S0 fraction with redshift and a simple scenario in which star
formation in the spiral progenitors of S0s was truncated at some time in
the past. More complex star formation histories or even ongoing star
formation in S0s may be the explanation. An alternative interpretation
is revealed by the stellar mass and dynamical mass TFRs. The TFR offset
persists as a function of both stellar and dynamical mass, and we show
that this may be evidence of a small (10--20\%) but systematic
contraction of spirals as they transform to S0s. This is consistent with
the trend with morphological type of the size--luminosity relation in
the local universe \citep{Courteau:2007}. If, on the other hand, the
zero point of the TFR has dimmed from the present day to high redshift
by $\approx 0.5$\,mag, then the putative transformation from spirals to
S0s involves essentially no movement in the velocity--luminosity plane.

It seems clear that S0s are not primeval objects, but are an end point
of spiral evolution. The processes responsible for this evolution can
perhaps be accelerated by the environment. The variation of the zero
point of the TFR with galaxy type and other parameters is just one
approach among a number available that will allow us to constrain S0
formation and evolution. It is uniquely powerful because it encodes
information about the amount and distribution of dynamical mass of a
galaxy. Modelling work complementary to the observations presented here
is underway \citep{Trujillo-Gomez:2010,Tonini:2010}. One should be sure
that the inevitable compromises introduced by high redshift observations
do not introduce biases similar to those ones discussed in
\refsec{sec:rotation}. We should also not forget the crucial role that
spectroscopy has to play, especially in constraining the star formation
history. 

\section*{Acknowledgements}

It is a pleasure to thank Alejandro Bedregal, Alfonso Aragon-Salamanca,
Michael Merrifeld and Karen Masters for their generous comments and
suggestions, the referee for his careful and extremely useful report,
and Aeree Chung and Giuseppe Aronica for access to the kinematic and
photometric data on which this project depends. MJW is supported by a
European Southern Observatory Studentship, MB by the STFC rolling grant
`Astrophysics at Oxford' (PP/E001114/1) and MC by an STFC Advanced
Fellowship (PP/D005574/1). We acknowledge with thanks the HYPERLEDA
database (\url{http://leda.univ-lyon1.fr}) and the NASA/IPAC
Extragalactic Database (NED) which is operated by the Jet Propulsion
Laboratory, California Institute of Technology, under contract with the
National Aeronautics and Space Administration.

\appendix

\section{Definitions of dynamical mass in disc galaxies}
\label{sec:mass}

In \refsec{sec:data} we discussed two ways to define or characterize the
dynamical mass of a galaxy. The first is to multiply the $K_S$-band
luminosity $L$ by the dynamical mass-to-light ratio
$(M/L)_{K_S,\mathrm{nohalo}}$ (taken from \citetalias{Williams:2009} for
our galaxies). As we noted, this method is relatively straightforward,
and we adopt it as our fiducial method in the main body of this paper
(denoting it $M_\mathrm{dyn}$). This quantity does not have a precise
physical meaning, but here we show it is equal, to a good approximation,
to $2\,M_{1/2}$, where $M_{1/2}$ is the dynamical mass contained within
the three-dimensional half-light volume. 

We estimate $M_{1/2}$ by inverting the expression for the circular
velocity of the mass model at the $r_{1/2}$. For a spherical system
(i.e. flattening $q \equiv c/a = 1$) this gives
\begin{equation}
M_{1/2} = r_{1/2} \vc^2(r_{1/2}) / G,
\label{eqn:mhalf}
\end{equation}
where $\vc(r_{1/2})$ is the circular velocity at $r_{1/2}$. Note that
$r_{1/2}$ is \emph{not} the projected half-light (effective) radius,
\reff. For a wide range of radial profiles, $r_{1/2} \approx 1.33\,\reff$
\citep{Hernquist:1990,Ciotti:1991,Wolf:2010}. 

\begin{figure}
\includegraphics[width=8.4cm]{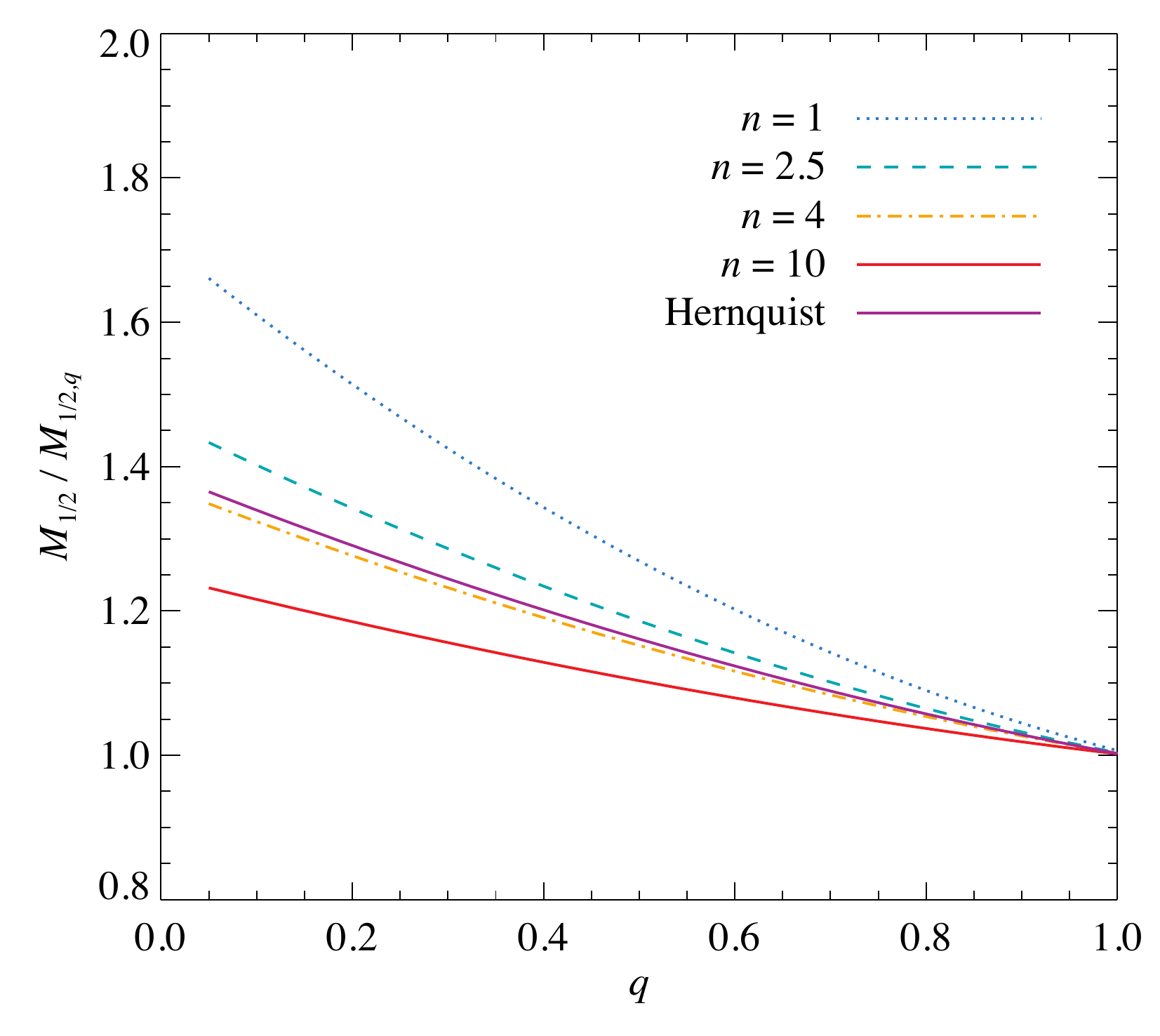}
\caption{$M_{1/2}/M_{1/2,\displaystyle q}$ for a selection of potentials as a function of
the flattening of the potential $q$. The behaviour of the
\protect\cite{Hernquist:1990} profile is shown in solid purple. The
other lines are S\'ersic profiles of order $n=1$ (exponential), 2.5 (typical
classical bulge--pseudobulge transition), 4 (de Vaucouleurs) and 10. The
de Vaucouleurs and Hernquist lines are well described by the fitting
formula $M_{1/2}/M_{1/2,\displaystyle q} = 1.38 - 0.52q + 0.14q^2$.}
\label{fig:flattening}
\end{figure}

Disc galaxies are not spherically symmetric, so a more useful definition
in these cases is the dynamical mass within an oblate spheroid of
semi-major axis $r_{1/2}$, which we denote $M_{1/2,\displaystyle q}$.
For a given circular velocity, $M_{1/2} > M_{1/2,\displaystyle q}$. This
is because the mass that must be spherically distributed to maintain a
given circular velocity is larger than that necessary to achieve the
same circular velocity in the midplane of a flattened system. The ratio
$M_{1/2}/M_{1/2,\displaystyle q}$ is simultaneously a function of the
radial profile of the mass distribution and $q$. For a full analytic
discussion see, for example, the comparison between the circular
velocities of oblate spheroids as a function of $q$ and mass profile
in section 2.5.2 of \cite{Binney:2008}. For our purposes, an approximate
estimate of $M_{1/2}/M_{1/2,\displaystyle q}$ is sufficient. Using a
range of analytic potentials, we examine the behaviour of the ratio
$M_{1/2}/M_{1/2,\displaystyle q}$ as a function of $q$ in
\reffig{fig:flattening}. The circular velocity of each flattened model
is conveniently calculated numerically by flattening a multi-Gaussian
fit to the mass distribution and using standard integrals to determine
the circular velocity of the arbitrarily flattened distribution
\citep{Cappellari:2002}. Disc galaxies typically have $q \approx 0.2$
and potentials of the form $\exp{(-R/\reff)^{1/n}}$, where $1 \le n \le
4$. In such galaxies, $1.3 \lesssim M_{1/2}/M_{1/2,\displaystyle q}
\lesssim 1.5$, and so from \refeq{eqn:mhalf}
\begin{equation}
M_{1/2,\displaystyle q} \approx r_{1/2} \vc^2(r_{1/2}) / 1.4 G.
\end{equation}
Furthermore, $r_{1/2} \approx 1.33\reff$, where in non-spherical systems the
observable \reff{} is defined as the semi-major axis of the projected
half-light ellipse. For disc galaxies we can therefore write
\begin{equation}
M_{1/2,\displaystyle q} \approx \reff \vc^2(1.33\,\reff) / G.
\label{eqn:newmhalf}
\end{equation}

The circular velocity at 1.33\,\reff{} can easily be estimated to an
accuracy consistent with the derivation of this formula from dynamical
modelling, stellar kinematics corrected for asymmetric drift or resolved
gas kinematic observations. For typical values of $q$ and the analytic
potentials explored here, this expression is accurate to about 15 per
cent. Of course in real disc galaxies, the flattening $q$ is a function
of radius, the potential is not described by a single S\'ersic or
Hernquist profile and even \reff{} is not simple to measure, so
\refeq{eqn:newmhalf} is
probably a little less reliable in practice. 

\begin{figure}
\includegraphics[width=8.4cm]{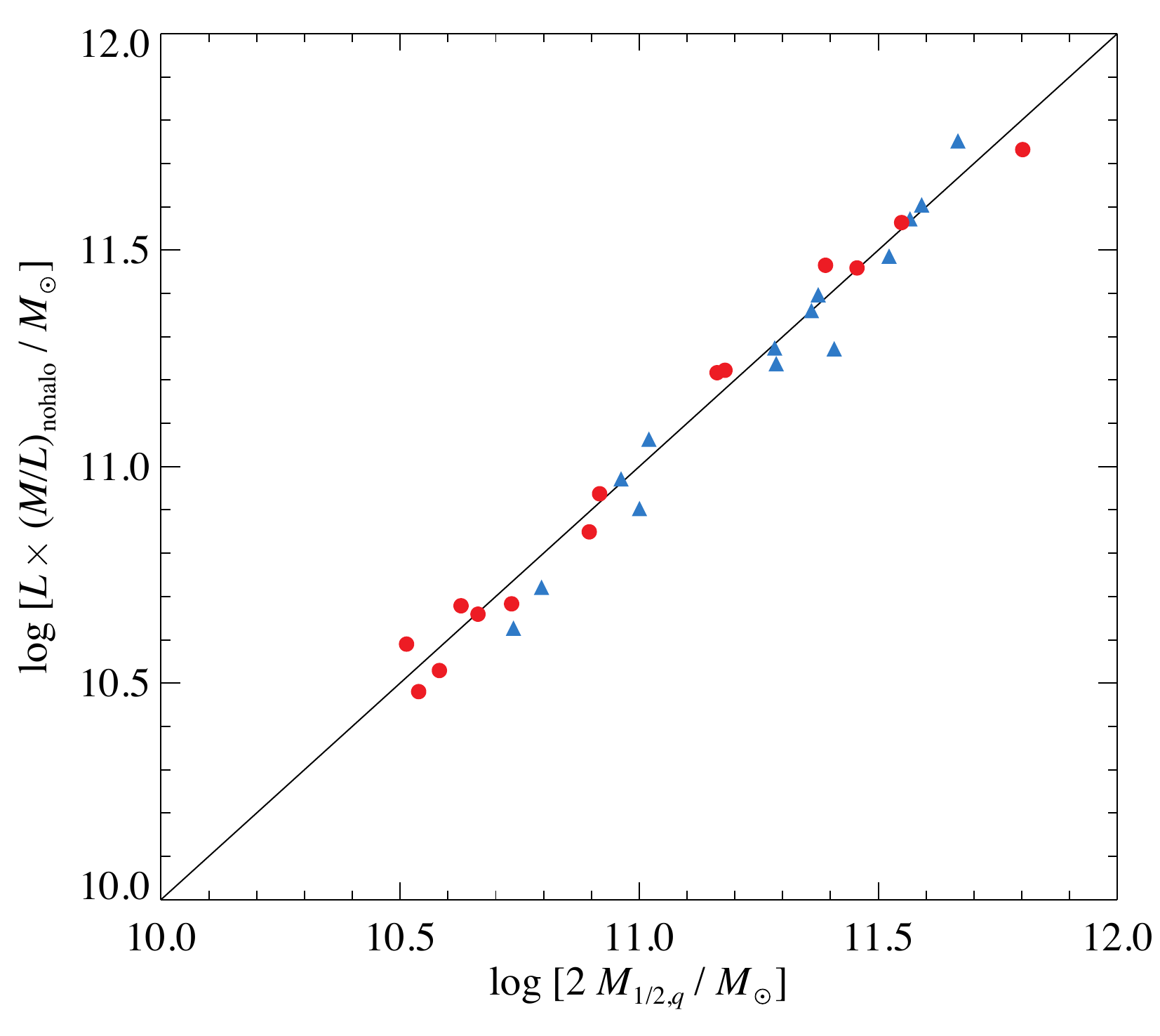}
\caption{A comparison of the two estimates of the dynamical mass for the
28 galaxies in the present sample. The horizontal axis is $2M_{1/2,\displaystyle q}$ as
defined in \refeq{eqn:newmhalf}, incorporating an approximate
correction for flattening. The vertical axis is the $K_S$-band
luminosity $L$ multiplied by the dynamical mass-to-light ratio
$(M/L)_{K_S,\mathrm{nohalo}}$ as presented in \citetalias{Williams:2009}. The symbols are as
in \reffig{fig:vall}. The solid line is the line $y = x$. The two
estimates agree on average to better than 2 per cent with 13 per cent
rms scatter and no evidence of a systematic difference between spirals
and S0s.}

\label{fig:comparemass}
\end{figure}

Nevertheless, when doubled, $M_{1/2,\displaystyle q}$ closely matches
fiducial estimates of the dynamical mass of our sample galaxies, i.e. the
product of their luminosities $L$ and dynamical mass-to-light ratios
$(M/L)_\mathrm{nohalo}$, inferred in this case from stellar kinematic
data extending to 2--3\,\reff{} and $K_S$-band photometry in
\citetalias{Williams:2009}. We
show this comparison in \reffig{fig:comparemass}. We use \vmodel{}, the
circular velocity of the \citetalias{Williams:2009} mass models in the flat region of the
observed rotation curve, as $\vc$ in \refeq{eqn:newmhalf}, but in cases
where a reliable asymmetric drift correction is possible at
1.33\,\reff{}, \vdrift{} gives almost identical results. The excellent
agreement between $2M_{1/2,\displaystyle q}$ and $L \times
(M/L)_\mathrm{nohalo}$ may be a coincidence. The two measures would be
systematically offset (and perhaps more scattered) if our kinematic data
extended significantly further in radius (and the dynamical
mass-to-light ratios we measured were therefore larger). Crucially,
however, \reffig{fig:comparemass} demonstrates that, for the purposes of
the present work comparing the TFRs of spirals and S0s, the choice of
definition of dynamical mass does not affect spirals and S0s
differently.
\label{lastpage}

\end{document}